\documentclass{aa}

\usepackage{amsmath}
\usepackage{natbib}
\usepackage{epsfig}
\usepackage{txfonts}

\newcommand{\ye}{Y_{\rm e}} 
\newcommand{\nue}{\nu_{\rm e}} 
\newcommand{\nuebar}{\bar \nu_{\rm e}} 
\newcommand{\Msol}{M_{\odot}}
\newcommand{\eexp}{{E_{\rm exp}}}

\newcommand{\amra}{A{\textsc{mra}}}

\bibpunct[, ]{(}{)}{;}{a}{}{,}

\begin{document}

\bibliographystyle{aa}

\title{Non-spherical core collapse supernovae}

\subtitle{II.~The late-time evolution of globally
              anisotropic neutrino-driven explosions
              and their implications for SN 1987\,A}

\author{
        K. Kifonidis\inst{1}  \and
        T. Plewa\inst{2}      \and
        L. Scheck\inst{1}     \and
        H.-Th. Janka\inst{1}  \and
        E. M\"uller\inst{1}
       }
        
\offprints{K. Kifonidis} 
\mail{kok@mpa-garching.mpg.de}

\date{received; accepted}      
       
\institute{Max-Planck-Institut f\"ur Astrophysik,
           Karl-Schwarzschild-Stra{\ss}e 1, 
           D-85741 Garching, Germany \and
           Center for Astrophysical Thermonuclear Flashes,
           University of Chicago, 5640 S. Ellis Avenue, 
           Chicago, IL 60637, USA }

\abstract{Two-dimensional simulations of strongly anisotropic
          supernova explosions of a nonrotating $15\,\Msol$ blue
          supergiant progenitor are presented, which follow the
          hydrodynamic evolution from times shortly after shock
          formation until hours later. It is shown that explosions
          which around the time of shock revival are dominated by
          low-order unstable modes (i.e. by a superposition of the
          $l=2$ and $l=1$ modes, in which the former is strongest),
          are consistent with all major observational features of SN
          1987\,A, in contrast to models which show high-order mode
          perturbations only and were published in earlier work. Among
          other items, the low-mode models exhibit final iron-group
          velocities of up to $\sim 3300$\,km/s, strong mixing at the
          He/H composition interface, with hydrogen being mixed
          downward in velocity space to only 500~km/s, and a final
          prolate anisotropy of the inner ejecta with a major to minor
          axis ratio of about 1.6. The success of low-mode explosions
          with an energy of about $2\times10^{51}$~erg to reproduce
          these observed features is based on two effects: the (by
          40\%) larger initial maximum velocities of metal-rich clumps
          compared to our high-mode models, and the initial global
          deformation of the shock. The first effect protects the
          (fastest) clumps from interacting with the strong reverse
          shock that forms below the He/H composition interface, by
          keeping their propagation timescale through the He-core
          shorter than the reverse shock formation time. This ensures
          that the outward motion of the clumps remains always
          subsonic, and that thus their energy dissipation is minimal
          (in contrast to the supersonic case).  The second effect is
          responsible for the strong inward mixing of hydrogen: The
          aspherical shock deposits large amounts of vorticity into
          the He/H interface layer at early times (around $t =
          100$\,s). This triggers the growth of a strong
          Richtmyer-Meshkov instability that results in a global
          anisotropy of the inner ejecta at late times (i.e. around $t
          = 10\,000$\,s), although the shock itself has long become
          spherical by then. The simulations suggest a coherent
          picture, which explains the observational data of SN 1987\,A
          within the framework of the neutrino-driven explosion
          mechanism using a minimal set of assumptions. It is
          therefore argued that other paradigms, which are based on
          (more) controversial physics, may not be required to explain
          this event.}

\maketitle  


\section{Introduction}
\label{sect:introduction}

Large-scale anisotropies and mixing, as well as smaller-scale
clumping, are common features of supernova (SN) remnants like Cas~A,
or SN 1987\,A, and have been attributed to the occurrence of
hydrodynamic instabilities (in particular the Rayleigh-Taylor or RT
instability) in the SN explosions that gave rise to these
remnants. Starting the supernova by depositing kinetic and/or thermal
energy into a progenitor model, many groups in the past have performed
multidimensional hydrodynamic simulations to study the mixing
associated with these effects. However, so far the models have not
been able to explain the observed features in a satisfactory way. An
extensive bibliography and overview of such work was given in the
first paper of the present series \citep[henceforth
Paper~I]{Kifonidis+03}.

In this latter reference we ourselves have embarked on a somewhat more
ambitious effort, namely to investigate the observational consequences
of the modern (i.e. multidimensional, or convectively supported)
neutrino-driven explosion paradigm for core collapse supernovae. For
this purpose we performed an exploratory, high-resolution 2D
simulation of a type~II supernova in a $15\,\Msol$ blue supergiant
which included a detailed modelling of the explosion itself, and which
we dubbed ``Model T310a''. This calculation covered the time from
20\,ms up to more than 5 hours after core bounce, and, at the time of
submission of the present paper, is still the only multidimensional
simulation that followed the hydrodynamic evolution and the growth of
anisotropies from the earliest moments after core collapse until well
after the time of shock emergence from the stellar photosphere. In
this way the effects of neutrino heating, convection, and explosive
nucleosynthesis deep inside the core, as well as their impact on the
Rayleigh-Taylor instabilities in the stellar envelope were all taken
into account.

For simulating the onset of the explosion in Paper~I, we approximated
the neutrino transport by a simple light-bulb scheme, and parametrized
the neutrino fluxes emitted by the optically thick proto-neutron star
core. In an attempt to reproduce the high nickel velocities of SN
1987\,A, the most sensitive test case for supernova theory available
to date, we performed the calculation using rather high initial values
for the neutrino fluxes, and a subsequent exponential decay of these
fluxes with time. This prescription initially appeared promising,
since it gave a rather energetic explosion. However, the maximum
nickel velocities that were finally obtained from the simulation, were
disappointingly small.

Furthermore, because the explosion in this model set in very rapidly,
it did not allow hydrodynamic instabilities, which occur in the
neutrino-heated post-shock layers, to grow to sufficient strength such
that they could deform the shock. Hence the model showed only
relatively small-scale deviations from spherical symmetry -- i.e.,
higher order modes of the fluid flow in terms of an expansion of the
inhomogeneities in Legendre polynomials of order $l$ -- and was not
able to reproduce the global anisotropy of the ejecta and the extent
of the mixing observed in SN 1987\,A. It also left behind a small
neutron star of only about $1.1\,\Msol$ (baryonic). Apparently, our
parametrization of the neutrino fluxes had resulted in too short an
explosion timescale, and our hope was that the aforementioned problems
might be overcome with an improved description of the neutrino
effects, e.g. by a modification of the (ad hoc) assumptions on the
behavior of the boundary conditions for the neutrino parameters.

This provided the motivation for the systematic two-dimensional
neutrino-hydrodynamic parameter study that we presented in
\cite{Scheck+04,Scheck+06}. In these works, we replaced the high
initial values of the core neutrino fluxes with lower ones, and the
(old) exponential decay with a \emph{much slower} decline over time,
an assumption which is in agreement with the results of current
simulations employing Boltzmann neutrino transport. (Note that the
core luminosity is imposed at our inner grid boundary which follows a
\emph{Lagrangian} mass coordinate.) Furthermore, we dropped the
simple neutrino light bulb approximation and replaced it with a new,
grey, characteristics-based neutrino transport scheme that we use in
the free-streaming and semi-transparent regimes. This allows us to
take into account the effects of neutrino heating and cooling on the
flux, e.g. to include the effect of mass accretion on the neutrino
luminosities (for details see \citealt{Scheck+06}).

With this new approach \cite{Scheck+04,Scheck+06} were able to
demonstrate that, if the explosion sets in sufficiently slowly,
low-mode hydrodynamic instabilities, i.e. low-mode convection
\citep{Chandra61,Herant95} in the neutrino-heating layer, and the
recently discovered advective-acoustic
\citep{Foglizzo_Tagger00,Foglizzo02,
Foglizzo_Galetti03,Foglizzo+05,Ohnishi+05}, and SASI instabilities
\citep{Blondin+03,Blondin_Mezzacappa05}, can grow from small random
perturbations behind the shock and result in a global anisotropy of
the shock and the ejecta. Moreover such explosions with typical
supernova energies also leave behind neutron stars with reasonable
masses and high recoil velocities, all at the same time.
\cite{Scheck+06} furthermore showed that the occurrence of different
low-order unstable modes ($l=1, l=2$) in the post-shock flow, might
even explain a possible bimodality of the observed galactic pulsar
velocity distribution without the need to assume any physics in
addition to what is already part of the neutrino-driven explosion
paradigm.

Yet, it needs to be pointed out that currently the viability of
neutrino-driven supernova explosions for progenitor stars of more than
$\sim 10\,\Msol$ is still an open question \citep[see,
e.g.,][]{Buras+03,Buras+06,Buras+06b}. Our adherence to the
neutrino-driven explosion mechanism and our procedure of triggering
supernova explosions by the use of a boundary condition for the
neutrino flux may be justified by the fact that uncertainties are
still present even in the most sophisticated recent 2D supernova
simulations, despite the significant improvements in the treatment of
neutrino physics and transport that have been achieved compared to the
first generation of multidimensional modelling. Such recent 2D
simulations, which include \emph{spectral} neutrino transport,
\citep[see][]{Buras+03,Buras+06,Buras+06b} could not confirm the
success of first generation models, which employed gray neutrino
diffusion and found explosions
\citep[e.g.][]{HBFC94,BHF95,Fryer99,FW02,Fryer+04}. 

However, it should be noted that 3D simulations with spectral
transport have not been performed yet, and might reveal important
differences in the flow dynamics compared to axisymmetric models, in
particular with respect to the growth of convective and
Rayleigh-Taylor instabilities, non-radial accretion shock
instabilities, and vortex behaviour.  Moreover, the question whether
convective or mixing processes below and around the neutrinosphere
could enhance the neutrinospheric neutrino emission and thus support
the neutrino heating behind the shock, must still be considered as
unsettled: Doubly diffusive instabilities \citep{Bruenn+06},
neutrino-bubble instabilities \citep{Socrates+05}, or magnetic
buoyancy instabilities \citep{Wilson+05} deserve further investigation
by multidimensional modelling. Also magneto-rotational instabilities
were suggested to aid the explosion by creating viscous heating behind
the shock in addition to the still dominant energy input there by
neutrinos \citep{Thompson+05}.

Although these currently unresolved issues need to be kept in mind,
the (2D) simulations of \cite{Scheck+06} suggest a quite remarkable
perspective, namely that many fundamental properties of observed
supernovae and neutron stars might be traced back to the same origin,
i.e. to non-radial hydrodynamic instabilities during the first second
of a neutrino-driven SN explosion. However, as these calculations did
not cover also the later phases of the explosion, they could not
provide insight into the inevitable interaction of the inner ejecta
with the (possibly massive, e.g. hydrogen) envelope of the
progenitor. Since this interaction (which happens on a timescale of
hours to days, depending on the type of the progenitor) determines
many of the more intricate observational features, especially of
Type~II supernovae, an important question still remains unanswered:
Can globally anisotropic neutrino-driven explosions also lead to
high final nickel velocities, strong mixing at the He/H composition
interface, large final ejecta anisotropies, and hence sizeable
polarization of the emitted light?

A positive answer to this question would not only offer a (natural)
explanation for the gamut of observational data collected in case of
SN 1987\,A, or SN remnants like Cas~A. It would also provide
substantial support to the neutrino mechanism. The present paper is an
attempt to explore this question in the framework of two-dimensional,
i.e. axisymmetric simulations. We exemplarily follow the evolution of
globally anisotropic explosions of the type studied in
\cite{Scheck+04,Scheck+06} from several milliseconds up to several
hours after core bounce in high-resolution simulations. 

Although we do not simulate the ``full SN problem'', but trigger the
neutrino-driven explosion by the use of a suitable inner boundary
condition, we nevertheless think that our modelling approach is
adequate to address the aforementioned points because it is
``consistent'' in the following sense: The distribution of kinetic and
internal energy in the flow, the number and size of flow structures
due to hydrodynamic instabilities, the shape and asymmetry of the
supernova shock, and the distribution of nucleosynthetic products
(both in real and velocity space) develop naturally in a detailed
neutrino-hydrodynamic calculation, in response to a boundary condition
for the neutrino flux and the contraction of the neutron star
core. Neither of the former quantities, which determine the kind of
observables that we are interested in, is put in ``by hand'' -- in
contrast to previous work in which ad hoc prescriptions were used for
at least one of these aspects.

The question of how close our boundary condition and initiation of the
explosion come to reality cannot be finally assessed at present,
because of the lack of detailed observed neutrino light curves, and/or
the aforementioned uncertainties inherent to more sophisticated
modelling of the (neutrino-driven) explosion mechanism. Yet, the
choice of particular parameter values for this boundary condition
affects mainly the integral characteristics of the explosion like its
total energy and time of onset, as well as the integral properties of
the neutron star, e.g. its mass and contraction time scale (see again
\citealt{Scheck+06} for details). A particular parametrization is in
the end vindicated by its success to reproduce the observational
features of SN 1987\,A.

The stellar model that we shall consider is the $15\,\Msol$ blue
supergiant SN 1987\,A progenitor of \cite{WPE88}, which we already
used in Paper~I. This makes the present work a direct sequel and
extension of this earlier investigation. It is our hope that these
papers, together with the constraints obtained from observations, will
serve to elucidate the still poorly understood hydrodynamics of SN
1987\,A, and help to disclose the true nature of this and similar
events.

We shall proceed with a description of the most recent modifications
to our computer codes in Sect.~\ref{sect:codes}. We then comment
briefly on our initial data and some computational aspects in
Sect.~\ref{sect:computational}. The results of our calculations are
presented in Sect.~\ref{sect:results_blue}, and compared with SN
1987\,A in Sect.~\ref{sect:SN_1987A}. Our conclusions can be found in
Sect.~\ref{sect:conclusions}. The appendix finally contains a summary
of the AUSM+ numerical flux function of \cite{Liou96} that is used in
one of our hydrodynamics codes.

\section{Modifications to the codes}
\label{sect:codes}

The computer programs that were used for the present calculations, are
the neutrino hydrodynamics code of \cite{Scheck+04,Scheck+06} and the
version of the adaptive mesh refinement hydrodynamics code \amra\ that
we described in Paper~I. Both use the direct Eulerian version of the
Piecewise Parabolic Method (PPM) of \cite{Colella_Woodward84} to
integrate the hydrodynamics equations. Several changes were made to
the former code, though. In particular we have improved its potential
to resolve shocks, by updating the hybrid Riemann solver algorithm
which we employed in \cite{Scheck+04,Scheck+06}. In these works we
have made use of the standard (in PPM) exact Riemann solver to evolve
all zones on the grid, except for those zones which are located in the
vicinity of strong (grid-aligned) shocks, for which the HLLE flux of
\cite{Einfeldt88} was employed (cf. Paper~I). After replacing the HLLE
flux by the AUSM+ flux of \cite{Liou96} (see also
Appendix~\ref{sect:ausmp}) we now achieve both, post-shock states
completely free of odd-even decoupling, and very sharp shocks with
only one or two interior points (while with the HLLE flux at least
3--4 points are required).

We have also completely revised the nucleosynthesis modules of this
code (see the end of this section for our motivation). For zones with
temperatures $1.5\times10^{9}~{\rm K} < T < T_{\rm max}^{\rm net}$ we
now solve a 32 species nuclear reaction network instead of the
modified $\alpha$-nuclei network described in Paper~I. The isotopes
considered in the 32 species network are listed in
Table~\ref{tab:nuclei}. Nuclear data and rates for 141 reactions in
total stem from the compilation of F.-K. Thielemann (see
\citealt{TNH96,Hoffman+99}, and the references therein). The ordinary
differential equations (ODEs) constituting the nuclear network are
solved with the variable-order Bader-Deuflhard semi-implicit, stiff
ODE integrator, which employs sub-stepping and Richardson
extrapolation to achieve very large effective (time) steps
\citep{Bader_Deuflhard83,Press+92,Timmes99}.

For $T > T_{\rm max}^{\rm net}$ a nuclear statistical equilibrium
(NSE) solution is computed: Given the temperature, $T$, density,
$\rho$, electron fraction, $Y_{\rm e}$, and the parameters $Z_i$
(charge number), $A_i$ (mass number), $B_i$ (binding energy) and
$G_i(T)$ (partition function) of species $i$, we iteratively solve
(with a globally convergent multidimensional Newton-Raphson scheme)
the Saha equations
\begin{equation}
Y_{i} = (\rho N_A)^{A_i-1} \frac{G_i}{2^{A_{i}}} A_{i}^{3/2}
          \left( \frac{2 \pi \hbar^2}{m_u k T} \right)^
          {\frac{3}{2}(A_i-1)} \exp{(B_i/kT)} 
          Y_{\rm n}^{(A_i-Z_i)} Y_{\rm p}^{Z_i}
\label{eq:saha}
\end{equation}
and the equations of charge and mass conservation
\begin{equation}
\sum_{i} Z_{i} Y_i = Y_{e}, \qquad \sum_{i} A_{i} Y_i = 1
\end{equation}
for the neutron and proton abundances, $Y_{\rm n}$ and $Y_{\rm p}$,
respectively. From these follow the abundances $Y_{i}$ of all other
species by virtue of Eq.~(\ref{eq:saha}).

\begin{table}
\begin{center}
\caption{Isotopes of the 32 species nuclear reaction network}
\label{tab:nuclei}
\begin{tabular}{cccccccc}
\hline

n & p & $\rm ^{4}He$ & $\rm ^{12}C$  &
$\rm ^{16}O$ & $\rm ^{20}Ne$ & $\rm ^{24}Mg$ & $\rm ^{28}Si$ \\

$\rm ^{32}S$  & $\rm ^{36}Ar$ & $\rm ^{40}Ca$ & $\rm ^{44}Ti$ &
$\rm ^{50}Ti$ & $\rm ^{48}Cr$ & $\rm ^{54}Cr$ & $\rm ^{55}Cr$ \\

$\rm ^{54}Mn$ & $\rm ^{55}Mn$ & $\rm ^{56}Mn$ & $\rm ^{52}Fe$ &
$\rm ^{54}Fe$ & $\rm ^{55}Fe$ & $\rm ^{56}Fe$ & $\rm ^{57}Fe$ \\

$\rm ^{58}Fe$ & $\rm ^{55}Co$ & $\rm ^{56}Co$ & $\rm ^{56}Ni$ &
$\rm ^{57}Ni$ & $\rm ^{58}Ni$ & $\rm ^{60}Ni$ & $\rm ^{60}Zn$ \\

\hline
\end{tabular}
\end{center}
\end{table}

Nuclear screening was included in the NSE solution according to
\cite{Hix_Thielemann96}, but is currently neglected in the nuclear
network solver. In order to get a perfect match between the numerical
solution of the nuclear network ODEs and the NSE solution we had to
set $T_{\rm max}^{\rm net} = 8\times10^{9}~{\rm K}$, i.e. we use the
nuclear network well into the medium-temperature NSE regime, where the
composition consists of neutrons, protons, and $\alpha$-particles, but
essentially no heavy nuclei. The resulting algorithm proved to be very
robust and shows no difficulties in making smooth transitions from
temperatures as high as $\sim 10^{11}~{\rm K}$ to less than
$10^{9}~{\rm K}$, and vice versa. In addition, it turned out to be
very efficient. Only a small number of (main) steps per burning zone
are necessary with the Bader-Deuflhard method to obtain a relative
integration accuracy of $10^{-4}$. The employed $T_{\rm max}^{\rm
net}$ turned out to be still low enough to avoid the reaction rates
becoming extremely large, and hence nuclear timescales becoming
extremely small\footnote{Despite our efficient solution of the
32-species network, the simulations still require about a factor of
five more computer time than without computing the nucleosynthesis.}.

The coupling of the nuclear burning algorithm with the hydrodynamics
was done in the same operator-split, ``co-processing'' approximation
that we described in Paper~I, i.e. energy and composition changes from
the 32 species solution are currently not fed back into the solution
of the hydrodynamics, which instead still relies on the smaller
4-species NSE solution tabulated in the equation of state of
\cite{JM96}. Work is in progress to extend the implicit nuclear
algorithm with an energy equation \citep[see][]{Mueller86}, and with
the leptonic equation of state of \cite{Timmes_Swesty00}, in order to
obtain a fully coupled treatment of the composition changes, nuclear
source terms, equation of state, and hydrodynamics in future
simulations.

We note here that our primary motivation for coding the present method
was its enhanced robustness and efficiency over the implementation of
the nuclear physics that we used in Paper~I. This enables us to handle
situations in which the previous algorithm (wich was based on the
``classical'' first-order Euler backward time discretization), either
experienced convergence problems or required too many nuclear time
steps to cover a hydrodynamic step. As a side effect, and due to the
large gain in efficiency provided by the Bader-Deuflhard integrator
\citep[see][]{Timmes99}, we can now also solve larger reaction
networks. However, improvements in the accuracy of the nuclear yields
which may thereby be achieved, need to be contrasted with the fact
that the relative yields of many isotopes within the iron group
(e.g. the $\rm ^{56}Ni/^{58}Fe$ ratio), depend \emph{extremely}
sensitively on the electron fraction, $Y_{\rm e}$, and thus require
also a more sophisticated, non-grey neutrino transport scheme for
their accurate computation than the neutrino treatment we employ
here (see also the discussion of these issues in Paper~I). For this
reason we will give only integrated abundances (or mass fractions) for
the iron group in this paper, instead of individual abundances for
single iron-group nuclei.

\begin{figure*}[tpb!]
\centering
\begin{tabular}{ccc}
\includegraphics[width=5.2cm]{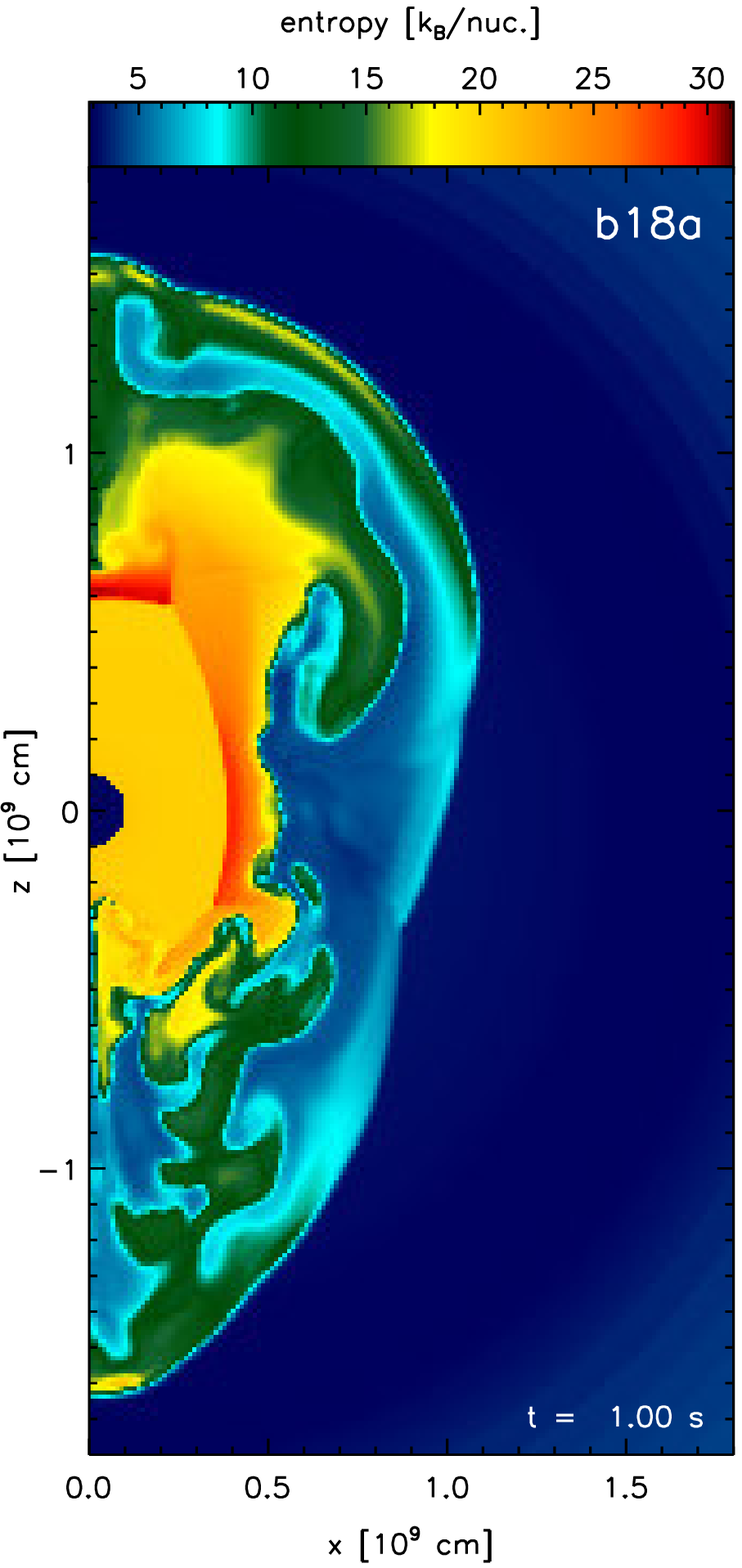} &
\includegraphics[width=5.2cm]{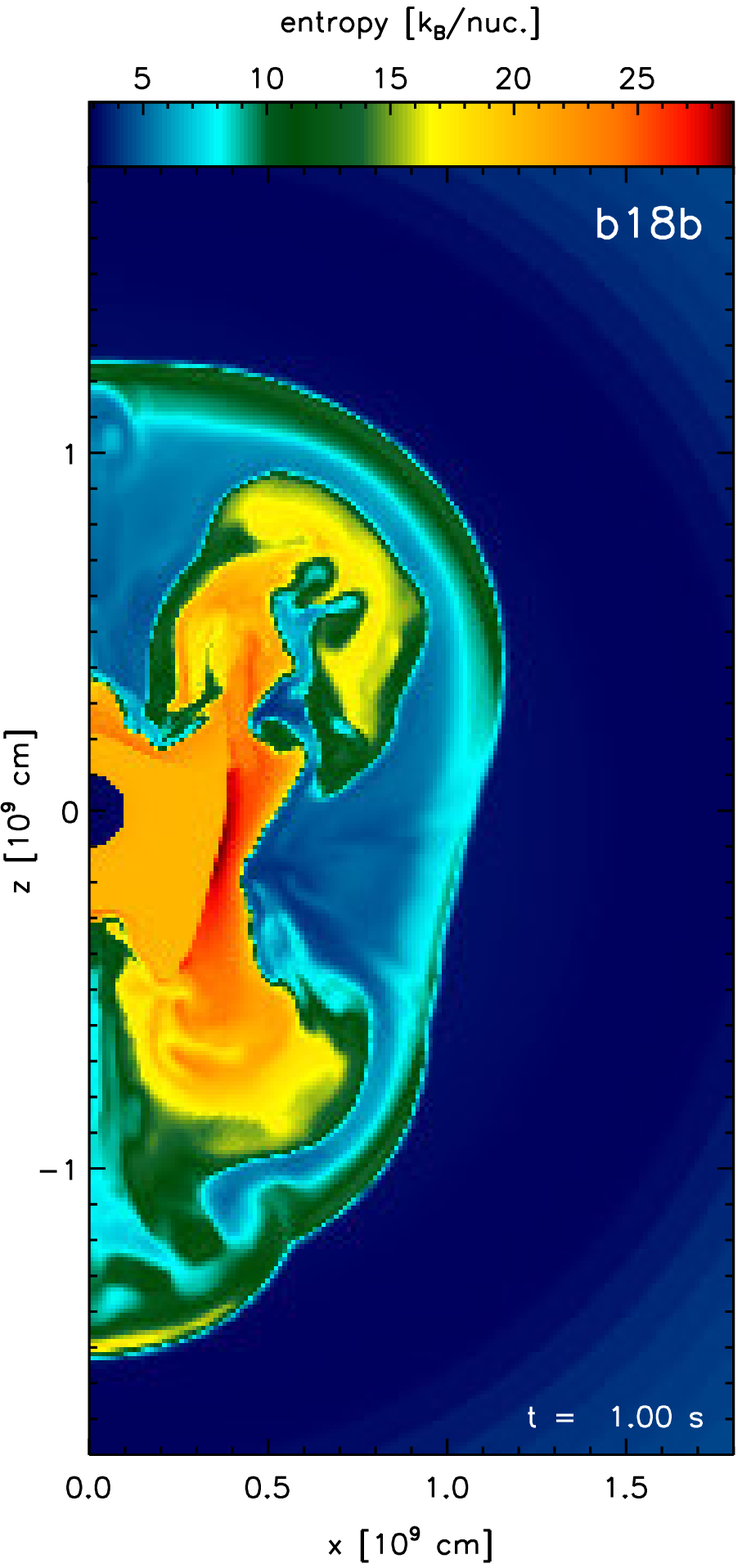} &
\includegraphics[width=5.2cm]{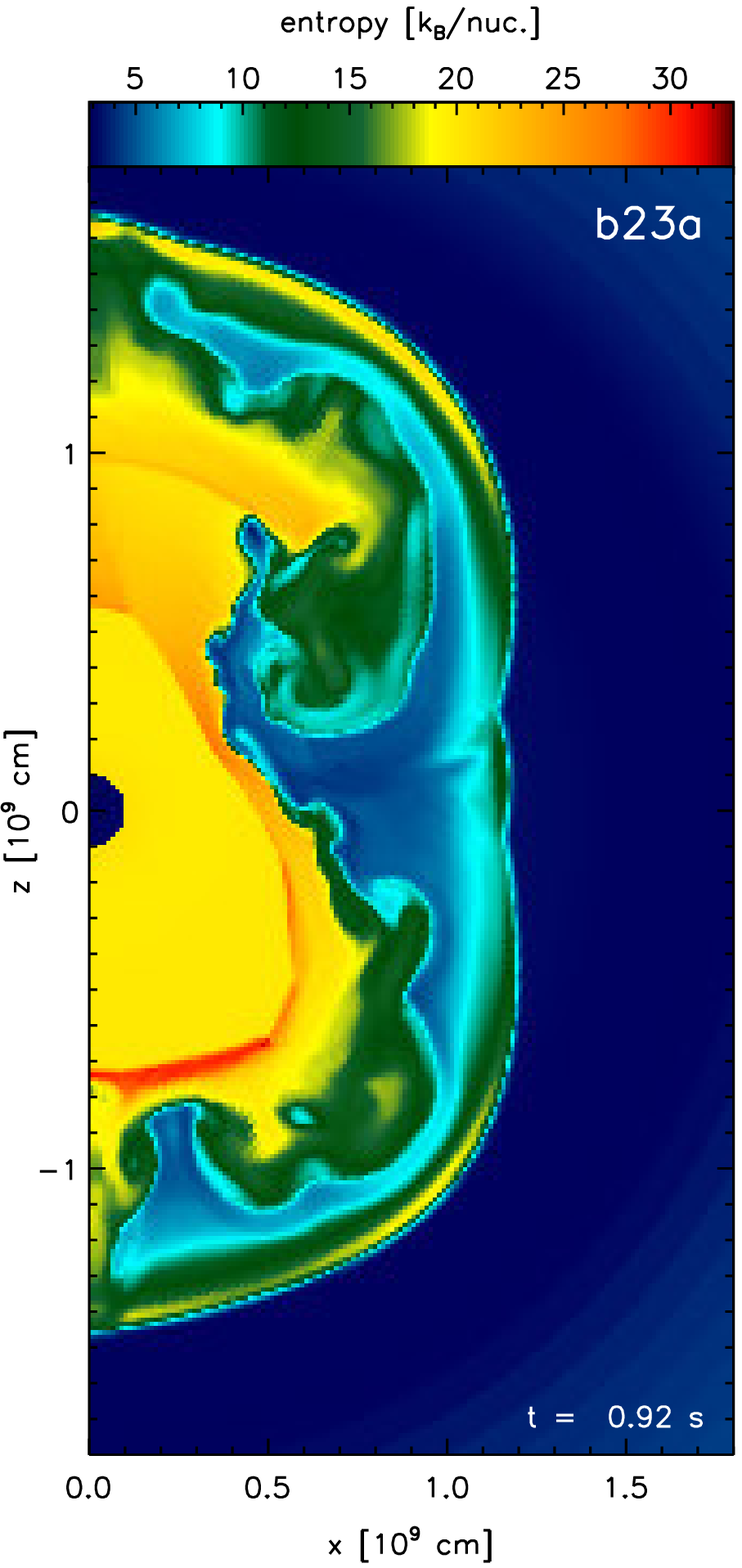} \\
\includegraphics[width=5.2cm]{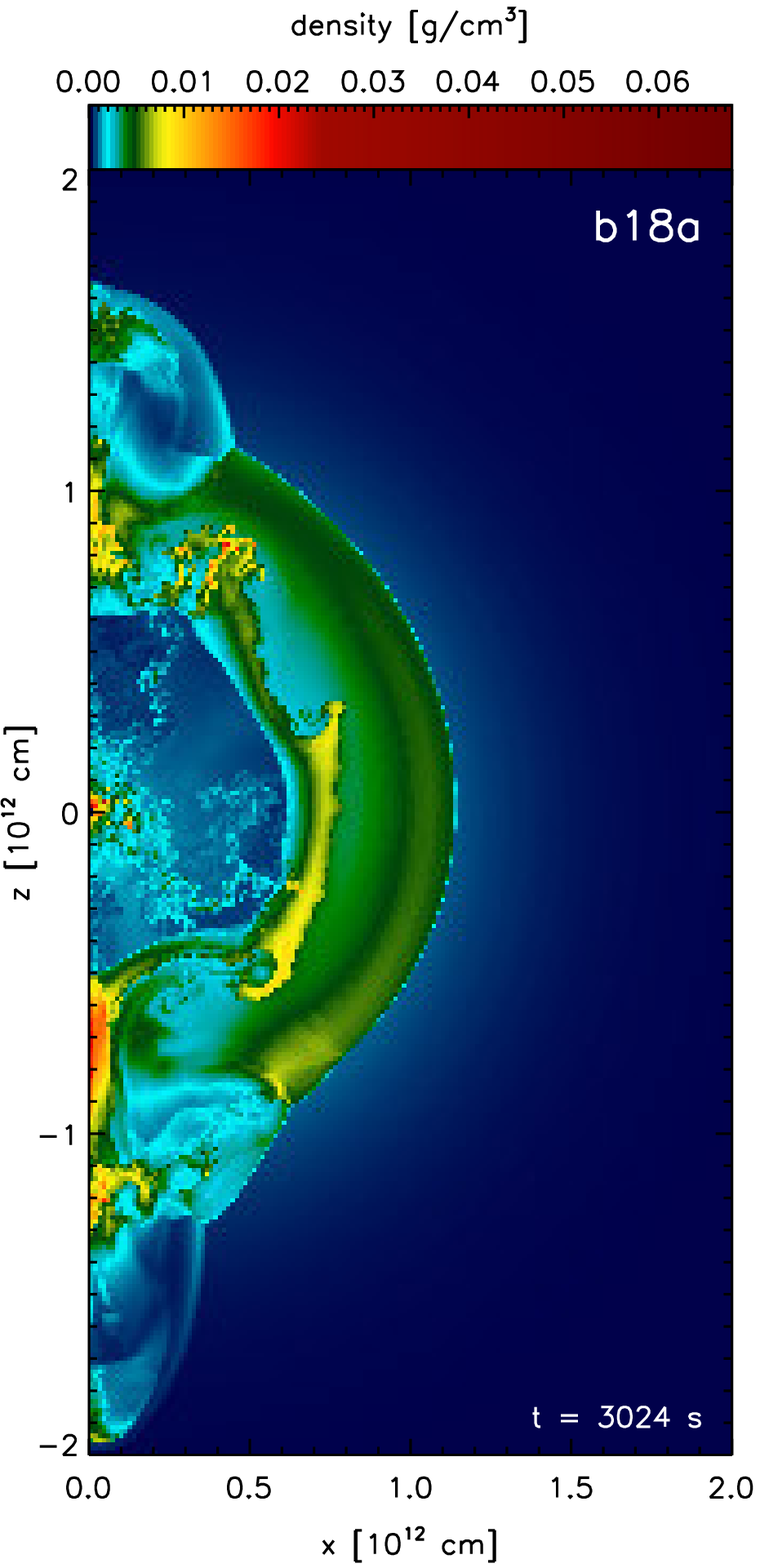} &
\includegraphics[width=5.2cm]{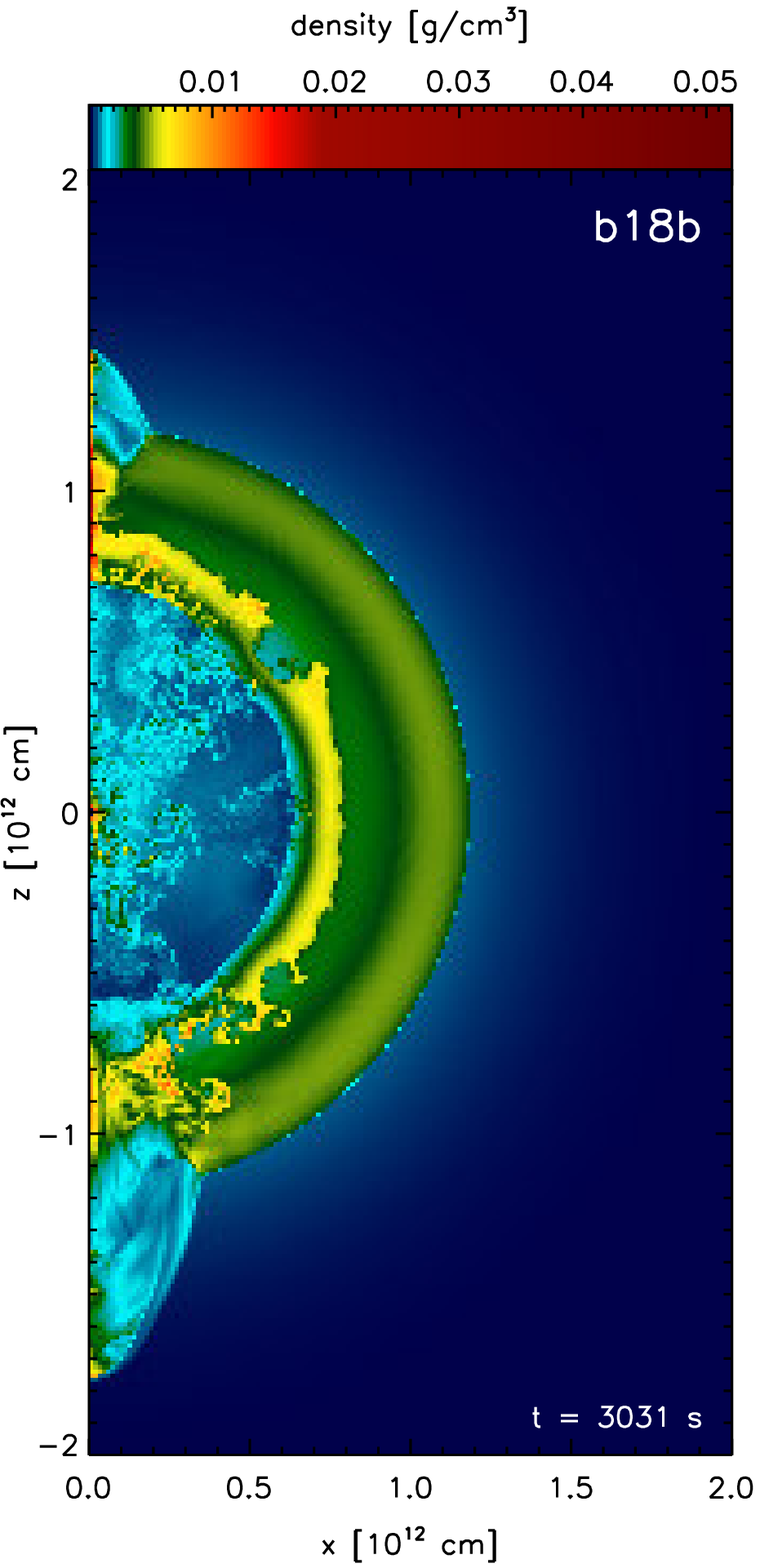} &
\includegraphics[width=5.2cm]{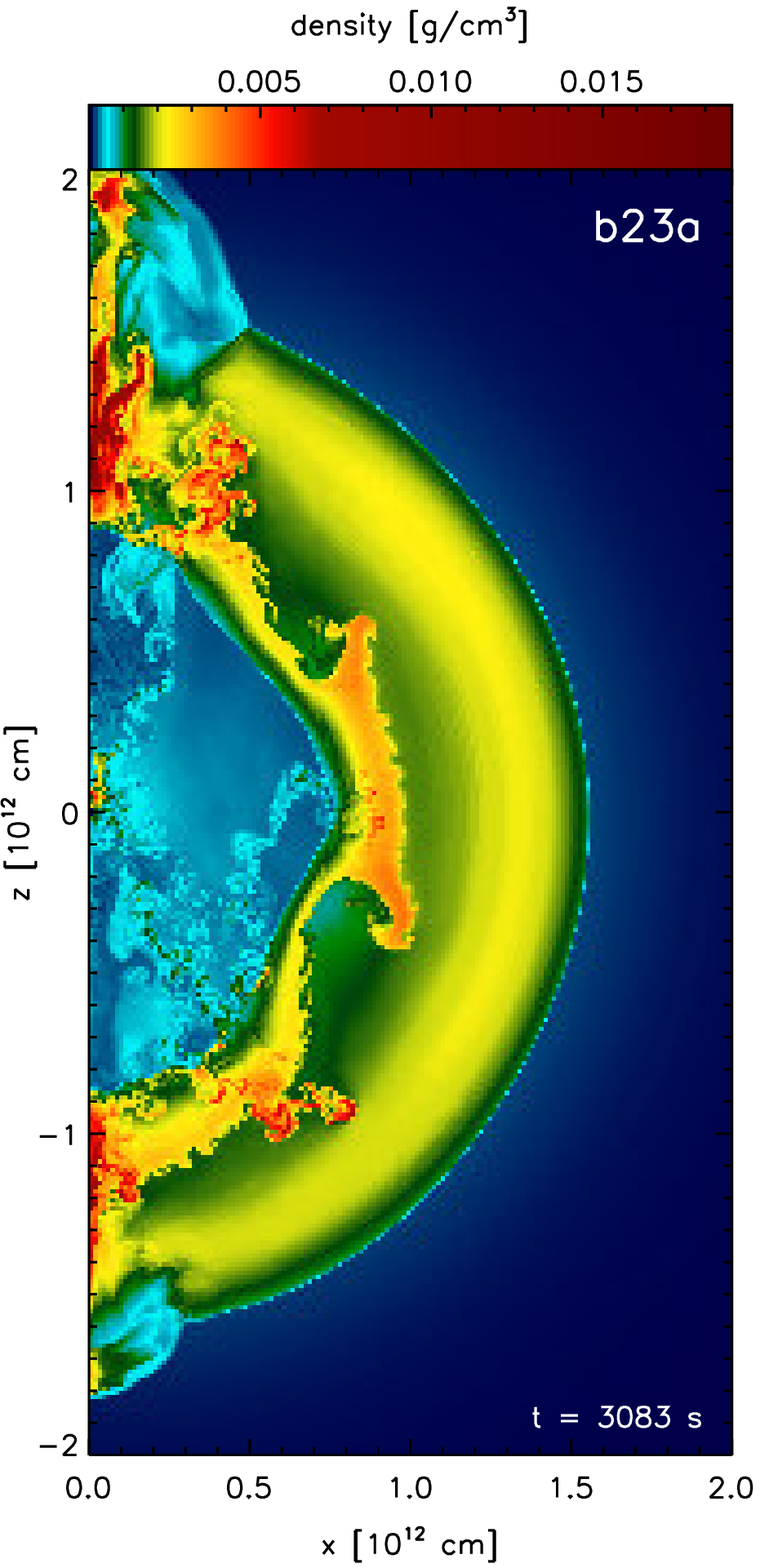} \\
\end{tabular}
\caption{Top row: Entropy distribution (in units of $k_{\rm
         B}$/nucleon) for three of the simulations listed in
         Table~\ref{tab:models_blue} at early times. From left to
         right a) Model b18a at $t=1$\,s, b) Model b18b at $t=1$\,s,
         and c) Model b23a at $t=0.92$\,s. Bottom row: Density
         distribution for the same models at later instants.  From
         left to right: d) Model b18a at $t=3024$\,s, e) Model b18b at
         $t=3031$\,s, and f) Model b23a at $t=3083$\,s. Note the
         change of shape of the supernova shock (the outermost
         discontinuity in all plots). Note also the polar jets which
         deform the otherwise spherical main shock at late times. They
         are a consequence of our use of a spherical coordinate grid
         and our assumption of axisymmetry.}
\label{fig:jets}
\end{figure*}

\section{Computational aspects}
\label{sect:computational}

\subsection{Numerical mesh, initial, and boundary conditions}
\label{sect:setup}

Throughout this work, 2D spherical coordinates $(r,\vartheta)$, a
computational grid with $180^{\circ}$ width in $\vartheta$, and
axisymmetry are adopted. To perform the present simulations we chose
to stick to the same $15\,\Msol$ blue supergiant model of \cite{WPE88}
and the post-bounce model of \cite{Bruenn93} (at a time of 20\,ms
after core bounce) that we have already used as initial data in
Paper~I.

Similar to this latter work, we follow the first second of the
explosion using neutrino-hydrodynamics calculations. These are set up
as described in \cite{Scheck+06} and we refer the reader to this paper
for details, in particular for the employed boundary conditions.

We wish to highlight one point though, which is connected to the fact
that we start our 2D simulations from supernova progenitor and
post-bounce models that are one-dimensional. Unlike the grid-less
smooth particle hydrodynamics (SPH) method, our grid-based Eulerian
hydrodynamics codes exhibit such a low level of numerical noise that a
one-dimensional, isotropic initial configuration remains perfectly
isotropic when evolved on a 2D grid, even if, e.g., a stratification
is present in the 1D initial data which is prone to become
convective. To trigger the growth of non-radial hydrodynamic
instabilities in the post-shock flow, we thus need to explicitly apply
a small random perturbation to the initial model, which we add to the
velocity field and for which we typically use an amplitude of
0.1\%. This perturbation is applied only once, at the beginning of the
calculation, i.e. at $t \approx 20$\,ms post-bounce.

To prescribe the neutrino radiation field of the contracting neutron
star core we make use of the same inner boundary contraction and
parametrization (with $\Delta E_{\rm core}^{\infty} = 0.18\, \Msol
c^2, t_L=1.0\,{\rm s}$) that was employed for Model B18 in
\cite{Scheck+06}. This resulted in a total electron flavour neutrino
luminosity of the neutron star core, $L^{\rm core} = L_{\nu_{\rm e}} +
L_{\bar \nu_{\rm e}} = 4.45 \times 10^{52}$~erg/s, which was assumed
to be constant during the first second.

In order to explore different explosion asymmetries resulting from the
non-radial instabilities, we calculated two models with this
particular setup that differ only in the initial perturbation. To
achieve this we simply used two different random number sequences to
perturb the initial velocity field in these two runs. Everything else,
in particular the perturbation amplitude that we employed, was kept
the same. The resulting models will henceforth be called b18a and
b18b. We point out here -- and elaborate on this in more detail in
\cite{Scheck+06} -- that already such minor differences between the
simulations produce different ejecta asymmetries, because the growth
of the initial perturbations is extremely nonlinear. In fact it is
chaotic.

We also used a second parametrization for the inner boundary condition
which differs from the former setup only in a somewhat larger
gravitational binding energy loss of the neutron star core, $\Delta
E_{\rm core}^{\infty} = 0.23\, \Msol c^2$. This leads to an $L^{\rm
core}$ which is by 28\% higher than in the b18 models, and gives rise
to Model b23a, which attained an explosion energy that is twice as
large as that of the b18 case (see Table~\ref{tab:models_blue} and
Sect.~\ref{sect:time_scales}).

Note that throughout this paper we use lower-case initial letters in
our model names to distinguish the present simulations from the models
of \cite{Scheck+06}, which did not make use of the nucleosynthesis
solver, and were still performed using the HLLE flux instead of the
AUSM+ flux (see above). Note also that in all of our simulations
(which were carried out with 400 radial and 180 angular zones), the
inner boundary condition for neutrinos was assumed to be
\emph{isotropic}, and that the core luminosities were held constant
over the time $t_L$ (for a motivation and full details, including a
comparison with results from Boltzmann transport calculations, see
\citealt{Scheck+06}).

The hydrodynamic evolution after the first second (which includes the
growth of Rayleigh-Taylor and Richtmyer-Meshkov instabilities) was
simulated with the \amra\ code and the mesh refinement hierarchy,
mesh resolution, remapping scheme, and boundary conditions
that were described in Paper~I. This allows us to temporarily
achieve a resolution equivalent to covering the entire progenitor with
$2.6\times10^{6}$ equidistant radial zones, although the effective
number of radial zones on the finest level of the AMR grid hierarchy
was only 3072. The effective number of angular zones on this
refinement level was 768.

\subsection{Constraints due to the grid-geometry}
\label{sect:grid_constraints}

At the outset we must acknowledge that a two-dimensional hydrodynamic
study in standard spherical coordinates has the drawback of messing
with the well-known coordinate singularity at the poles of the
spherical grid. From earlier calculations of the Rayleigh-Taylor
mixing in core collapse supernovae employing spherical coordinates it
is known (see \citealt{Kane+00}, and Paper~I) that Rayleigh-Taylor
fingers grow faster along the poles than in other regions of the grid,
and may even evolve to (artificial) ``axial jets'' if the simulation
is followed to late times. This effect is most likely enhanced by the
imposed axisymmetry, which restricts the degrees of freedom of the
flow, and promotes its convergence along the impenetrable poles.

If the initial conditions do not deviate too strongly from spherical
symmetry, the axial artifacts remain rather localized, and do not
affect the global character of the solution (see the calculations of
Paper~I, and \citealt{Kane+00}). However, this may no longer be the
case if one starts from very anisotropic situations. To illustrate
this, we show the distributions of entropy and density at $t \approx
1$\,s and at $t \approx 3000$\,s, respectively, for Models b18a, b18b,
and b23a in Fig.~\ref{fig:jets}. Of these models the first one (Model
b18a) is the most anisotropic, being dominated by the $l=1$ (bipolar)
mode one second after core bounce, while the latter two show a
dominance of the $l=2$ (quadrupole) mode (with a smaller $l=1$
contribution).

It is apparent from Figs.~\ref{fig:jets}d--f that in all of the models
the solution at a time of $\sim 3000$ seconds is affected
significantly in the polar regions, showing a pair of polar ``bulges''
whose origin is in all cases a pair of polar Rayleigh-Taylor
fingers. These expand with much higher velocity than the surrounding
material and ultimately deform the main shock.

Yet, while in Models b18b and b23a the solution is still undisturbed
at distances $\gtrsim 15-20^{\circ}$ from the poles, the artifacts are
very pronounced in the most anisotropic model, b18a. In this case they
have apparently also interacted with the \emph{physical} instability
discussed in Sect.~\ref{sect:Richtmyer}, affecting a cone as large as
$\sim 45^{\circ}$ around the south pole. We feel that the occurrence
of a possibly spurious solution over such a large fraction of the
computational domain makes this model unsuitable for reasonable
analysis, and we will thus skip it from the further discussion. Hence,
we will not follow the late-time evolution of this strongly one-sided
explosion (resulting from a clearly dominant $l=1$ mode) in this
paper, but will constrain ourselves to explosions in which the $l=2$
mode is dominant, and the $l=1$ mode yields only a smaller
contribution. We will also try to exclude the regions of these latter
models which are affected by the ``jets'' from all our analyses of
expansion velocities and the extent of mixing of different elements,
by skipping $15^{\circ}$ of the computational wedge closest to the
north and south poles from the data evaluation.

For similar reasons we will also not attempt to calculate the
late-time evolution of rotating models in this paper. It is actually
true that in a rotating model the polar axis, i.e. the axis of
rotation, represents a distinguished direction of the physical system,
which is impenetrable for the flow if the specific angular momentum of
fluid elements is conserved. In this case the hydrodynamic solution
may in fact yield the formation of jet-like polar outflows which are
physical. In 2D (i.e. axisymmetric) calculations, employing standard
spherical coordinates, it is, however, impossible to study the
formation of such features reliably, since the enforced conservation
of the $z$-component of the specific angular momentum leads to
numerical artifacts near the poles which will always contaminate the
numerical solution.

Unfortunately, a reliable modelling of late-time supernova
hydrodynamic instabilities (beyond times of $\sim 10$\,s) for
explosions with extreme initial anisotropies and/or rotation appears
to require high-resolution 3D simulations of the entire sphere, either
in Cartesian coordinates, or with a composite spherical mesh, as
e.g. the ``cubed sphere grid'' described in \cite{Ronchi+96}. Both
approaches are expected to result in computational tasks that are
between two and three orders of magnitude more expensive than the
present calculations.

\section{Results}
\label{sect:results_blue}

\begin{table}
\begin{center}
\caption{Models for the \protect \cite{WPE88} blue supergiant.  Given
         are the explosion timescale, $t_{\rm exp}$, and the entire
         simulated time, $t_{\rm sim}$, as well as the explosion
         energy at a time of 3000 seconds, $E^{t=3000{\rm s}}_{\rm
         exp}$, and the baryonic mass of the neutron star at $t
         \approx 1~{\rm s}$ (without fallback). The explosion
         timescale is defined by the time after core bounce when the
         integral of the total energy over all grid cells with
         positive values of that quantity and positive radial
         velocities, is larger than $10^{48}\,{\rm erg}$.}
\label{tab:models_blue}
\begin{tabular}{lccccc}
\hline
Model &$t_{\rm exp}$ &
              $E^{t=3000{\rm s}}_{\rm exp}$ & 
              $M_{\rm ns}$ &
              $t_{\rm sim} $ \\
 &  [s] & $[10^{51}\,{\rm erg}]$ & [$\Msol$] & $[{\rm s}]$ \\
\hline
b18a  (disregarded)  & 0.190 & 1.0 & 1.3 &$         10^4$  \\
b18b                 & 0.185 & 1.0 & 1.3 &$2 \times 10^4$  \\
b23a                 & 0.138 & 2.0 & 1.2 &$2 \times 10^4$  \\
\hline
T310a (Paper~I)      & 0.062 & 1.7 & 1.1 & $2 \times 10^4$  \\
\hline
\end{tabular}
\end{center}
\end{table}

\begin{figure*}[tpb!]
\centering
\begin{tabular}{ccc}
\includegraphics[width=5.2cm]{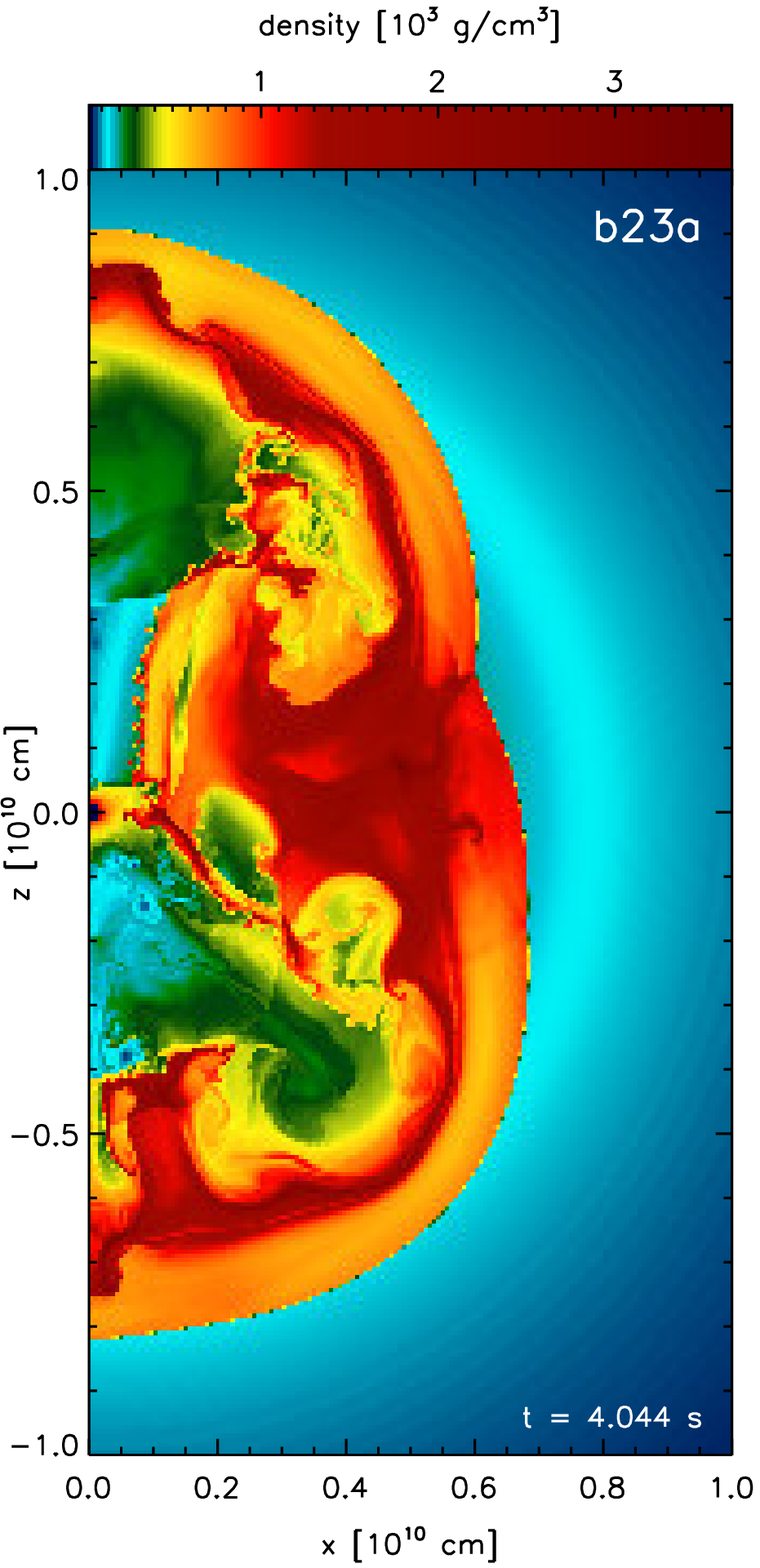}  &
\includegraphics[width=5.2cm]{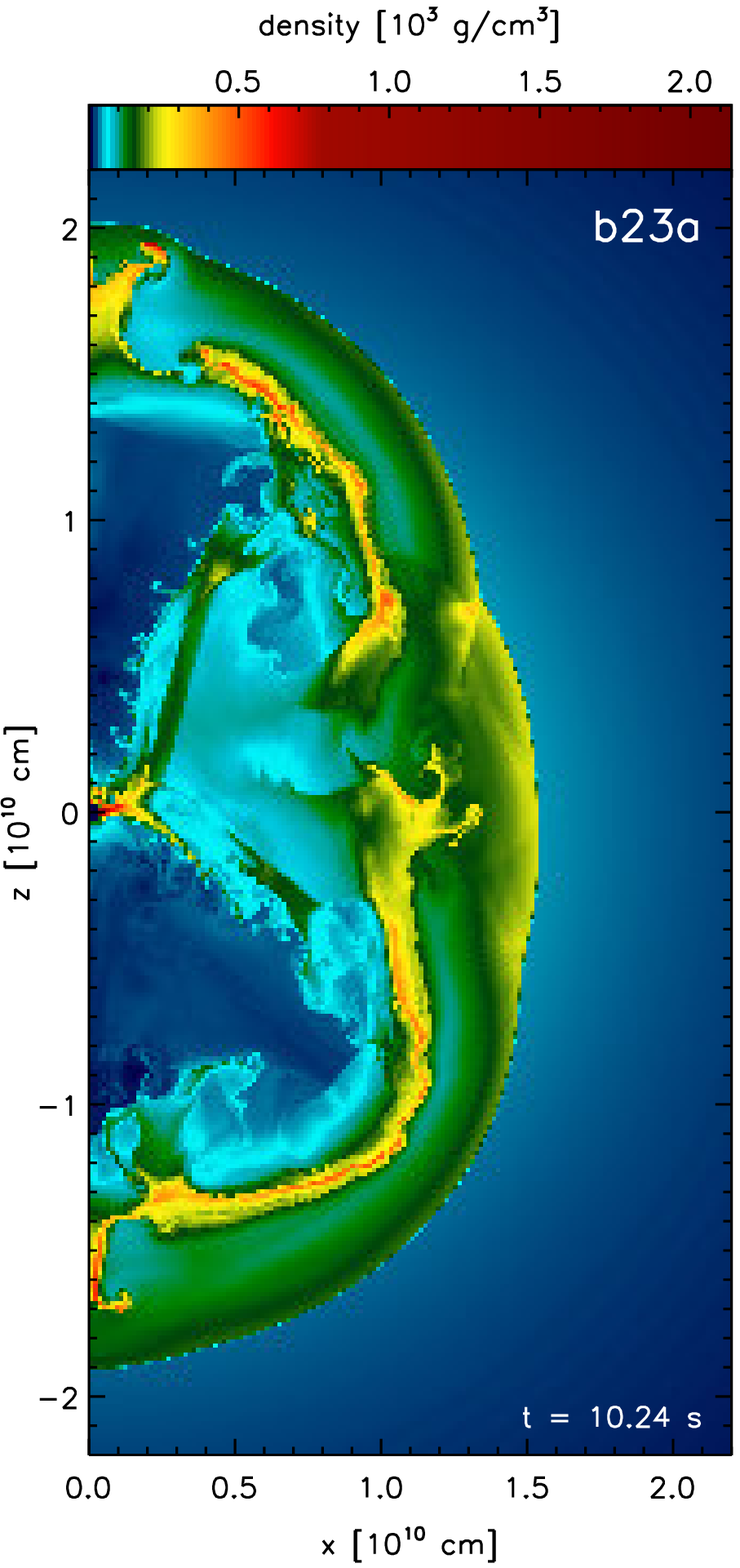} &
\includegraphics[width=5.2cm]{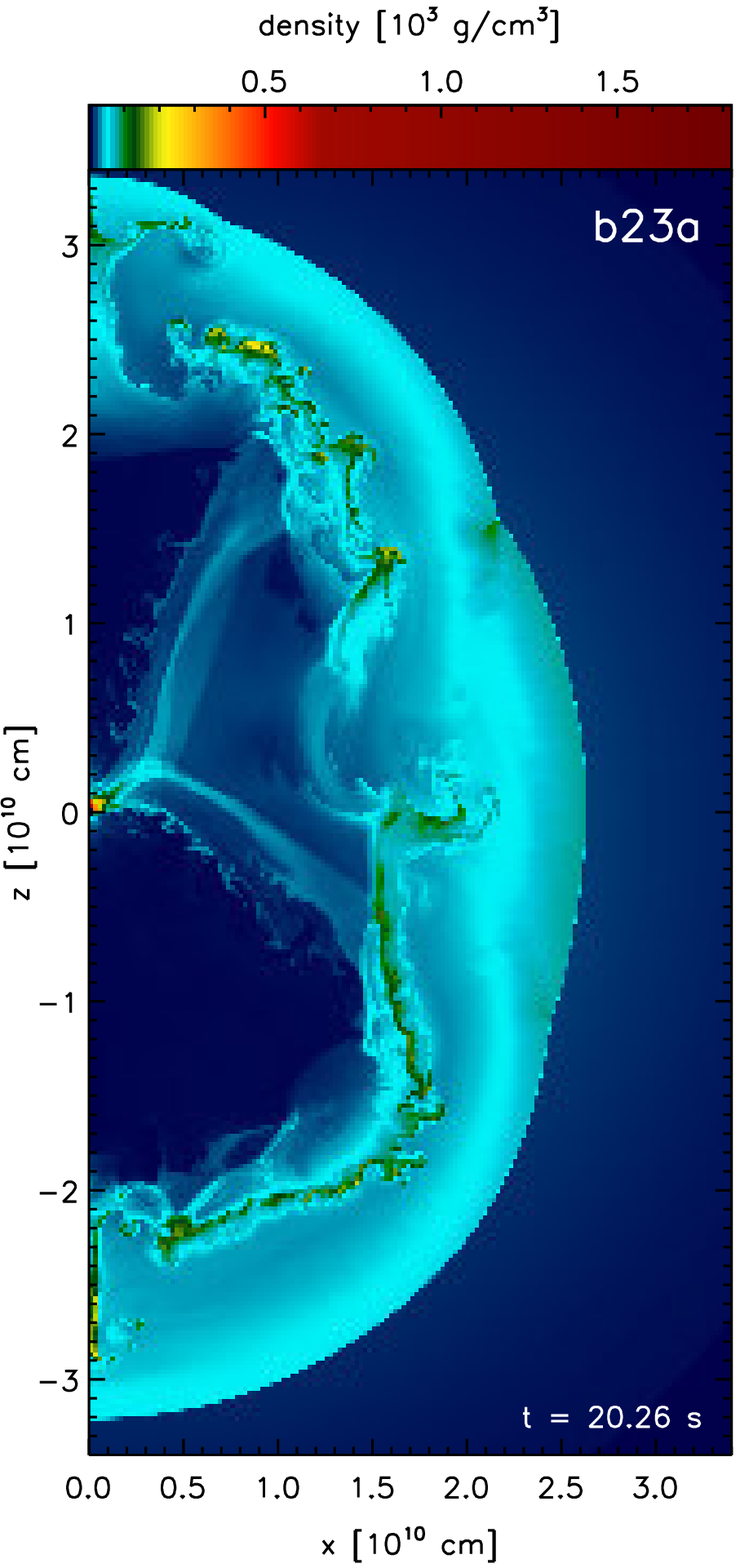} \\
\includegraphics[width=5.2cm]{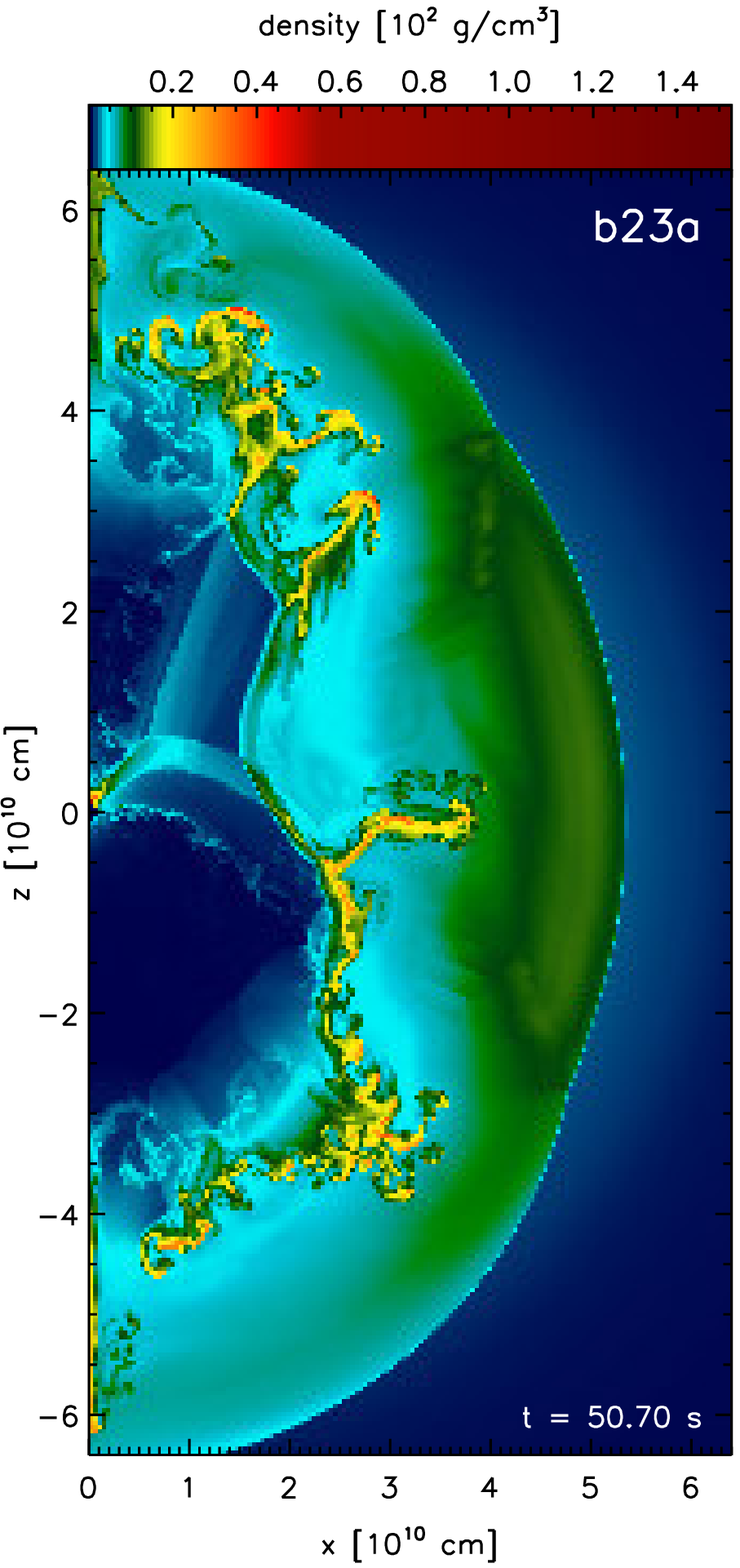}  &
\includegraphics[width=5.2cm]{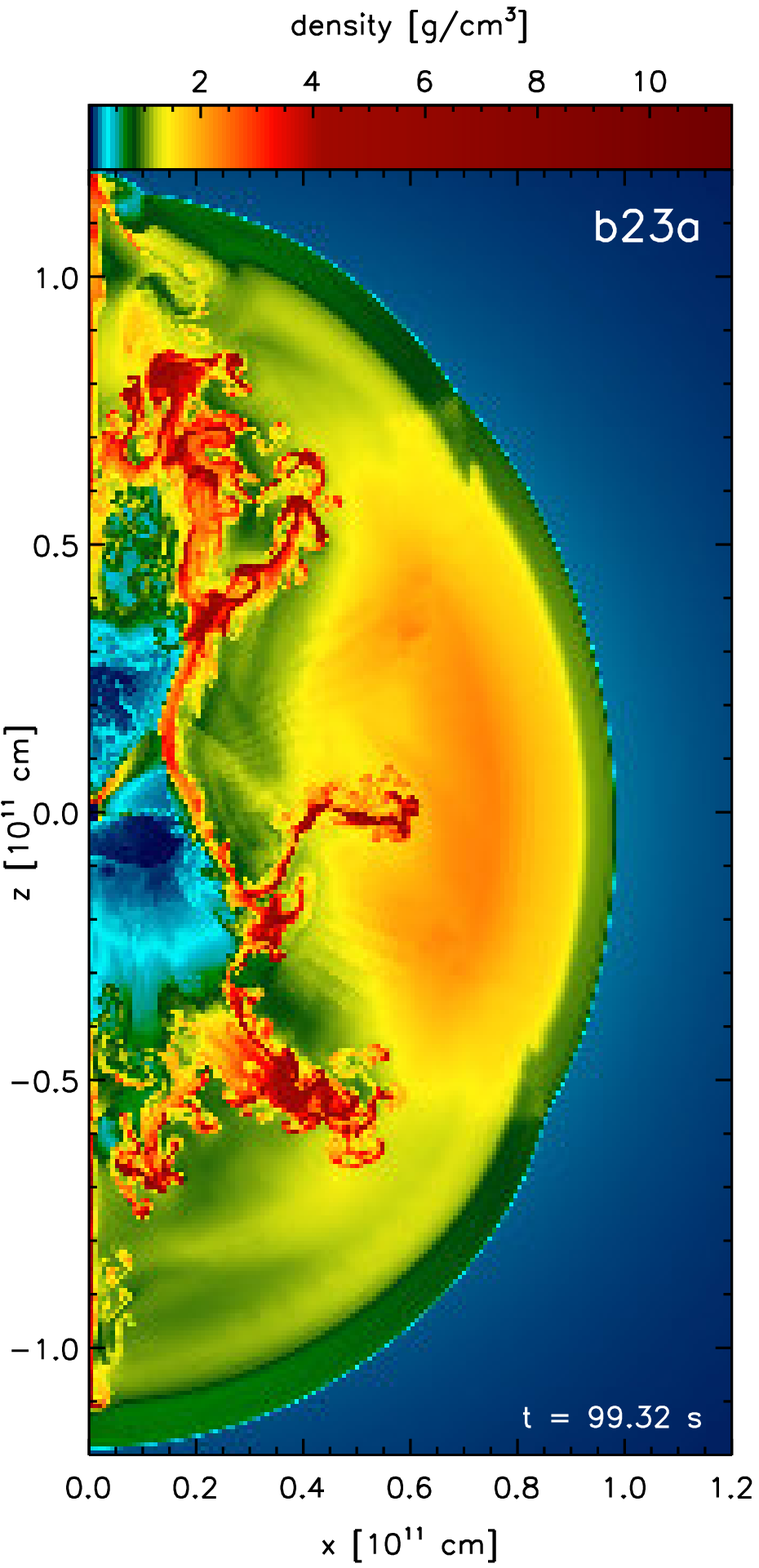} &
\includegraphics[width=5.2cm]{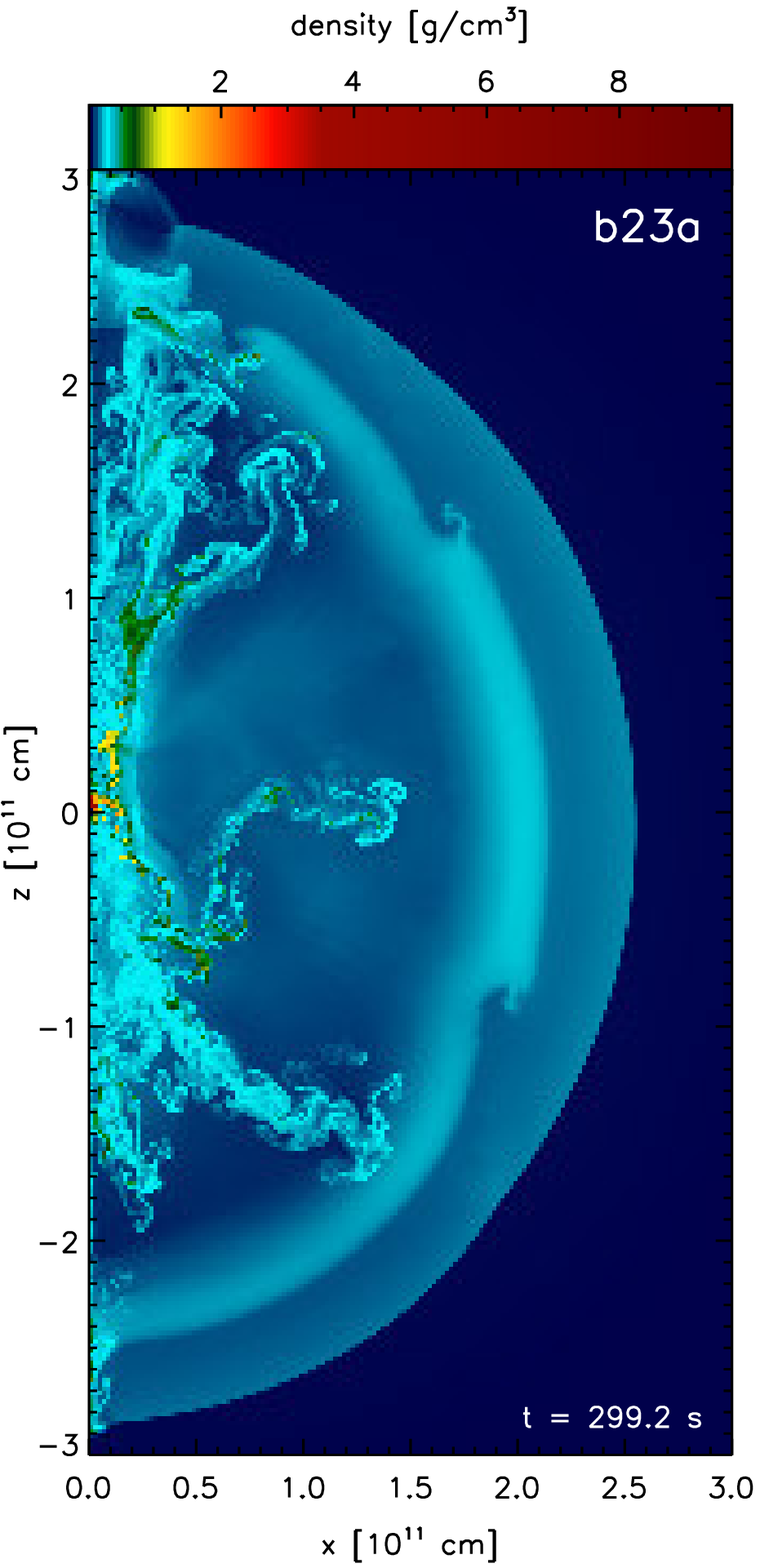}
\end{tabular}
\caption{Snapshots of the density distribution of Model b23a for
         several times. From top left to bottom right a) $t=4$\,s, b)
         $t=10$\,s, c) $t=20$\,s, d) $t=50$\,s, e) $t=99$\,s, and f)
         $t=299$\,s. Note the change of the radial scale. The
         supernova shock is the outermost discontinuity which becomes
         progressively spherical with time. Note also the growth and
         outward propagation of the Rayleigh-Taylor mushrooms from the
         former Si/O and O/He interfaces of the star, and the onset of
         the Richtmyer-Meshkov instability at the He/H interface (the
         deformed discontinuity just behind the supernova shock in the
         plots for $t=99$\,s and $t=299$\,s). }
\label{fig:b23a_dens_a}
\end{figure*}

\begin{figure*}[tpb!]
\centering
\begin{tabular}{ccc}
\includegraphics[width=5.2cm]{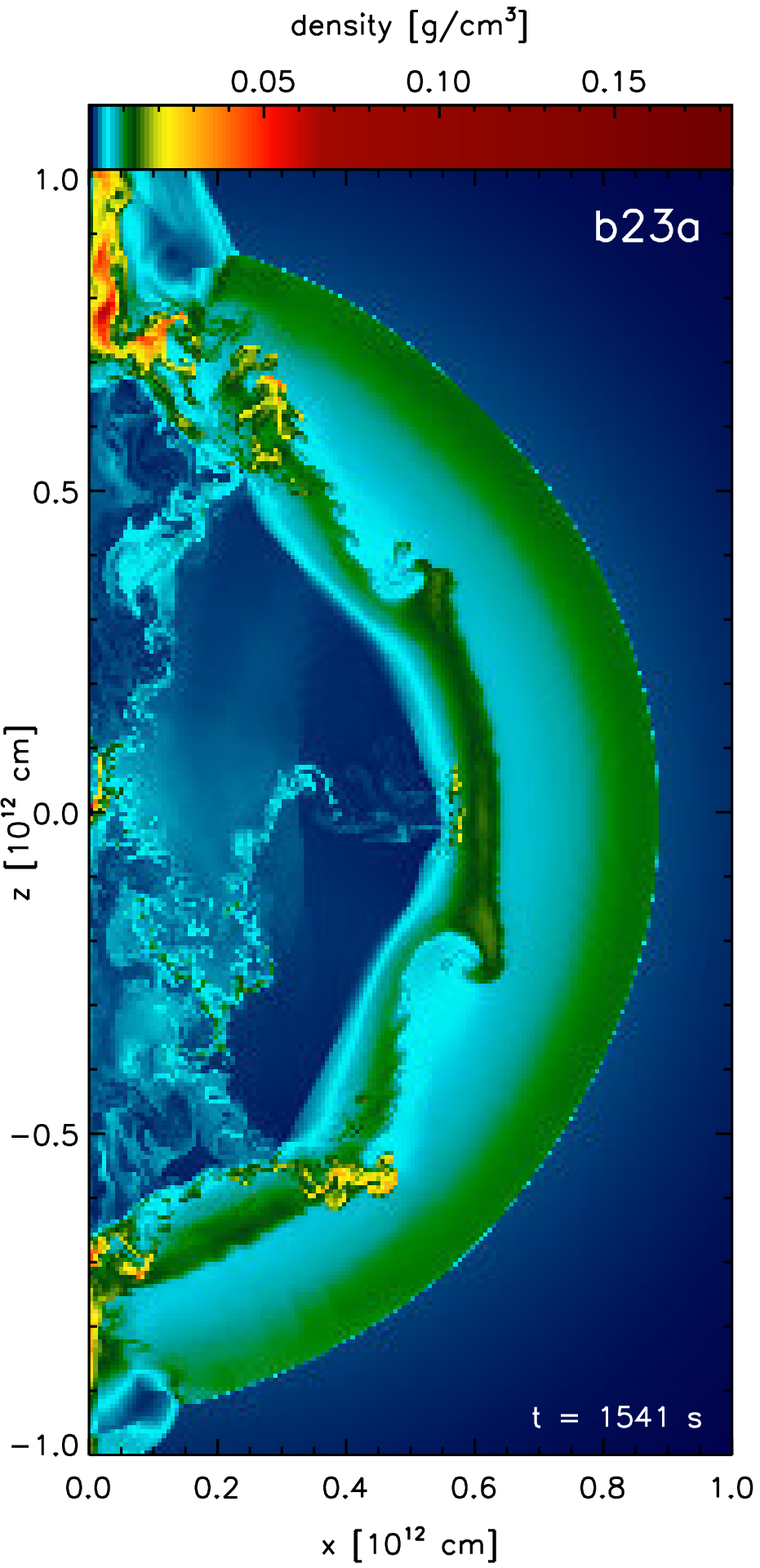}   &
\includegraphics[width=5.2cm]{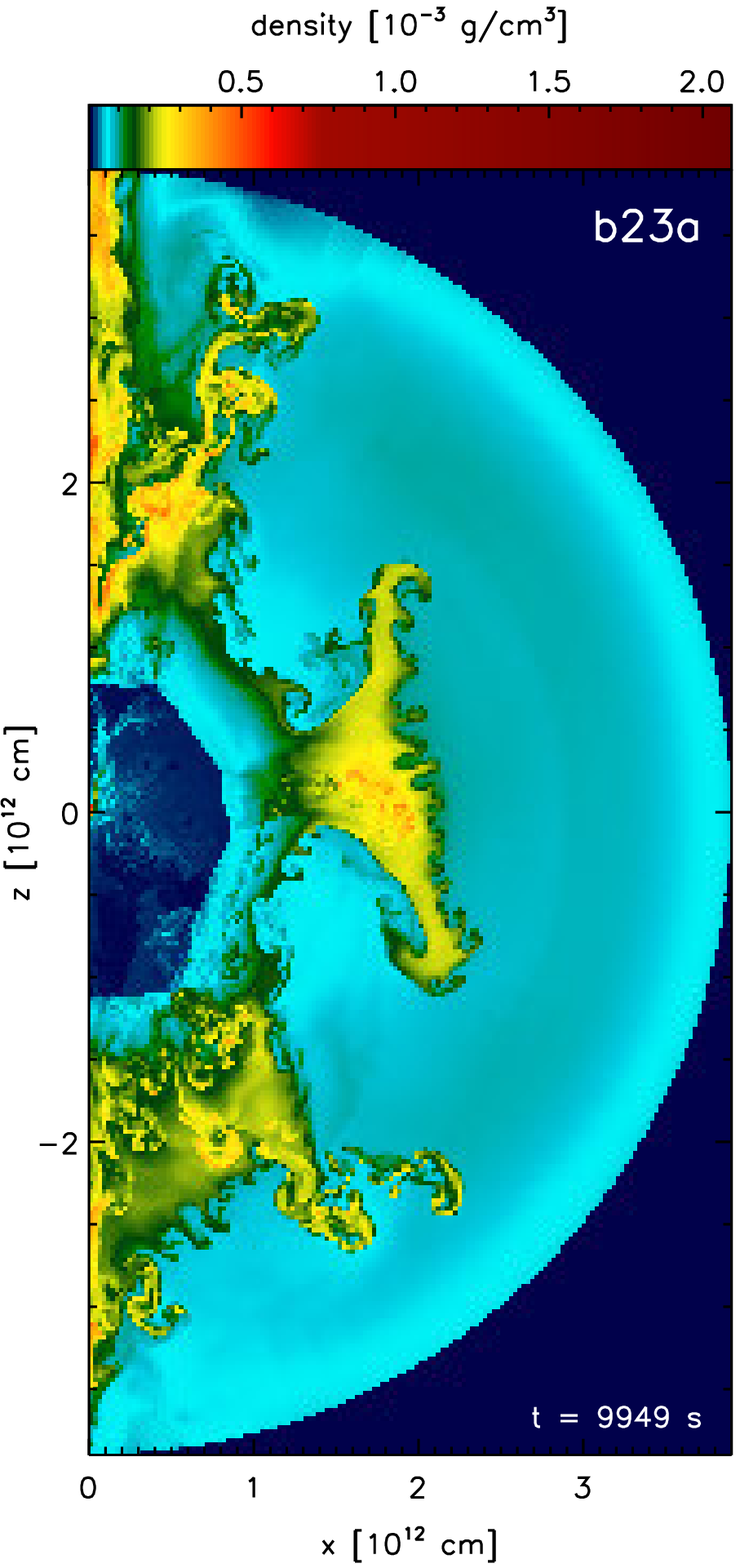}   &
\includegraphics[width=5.2cm]{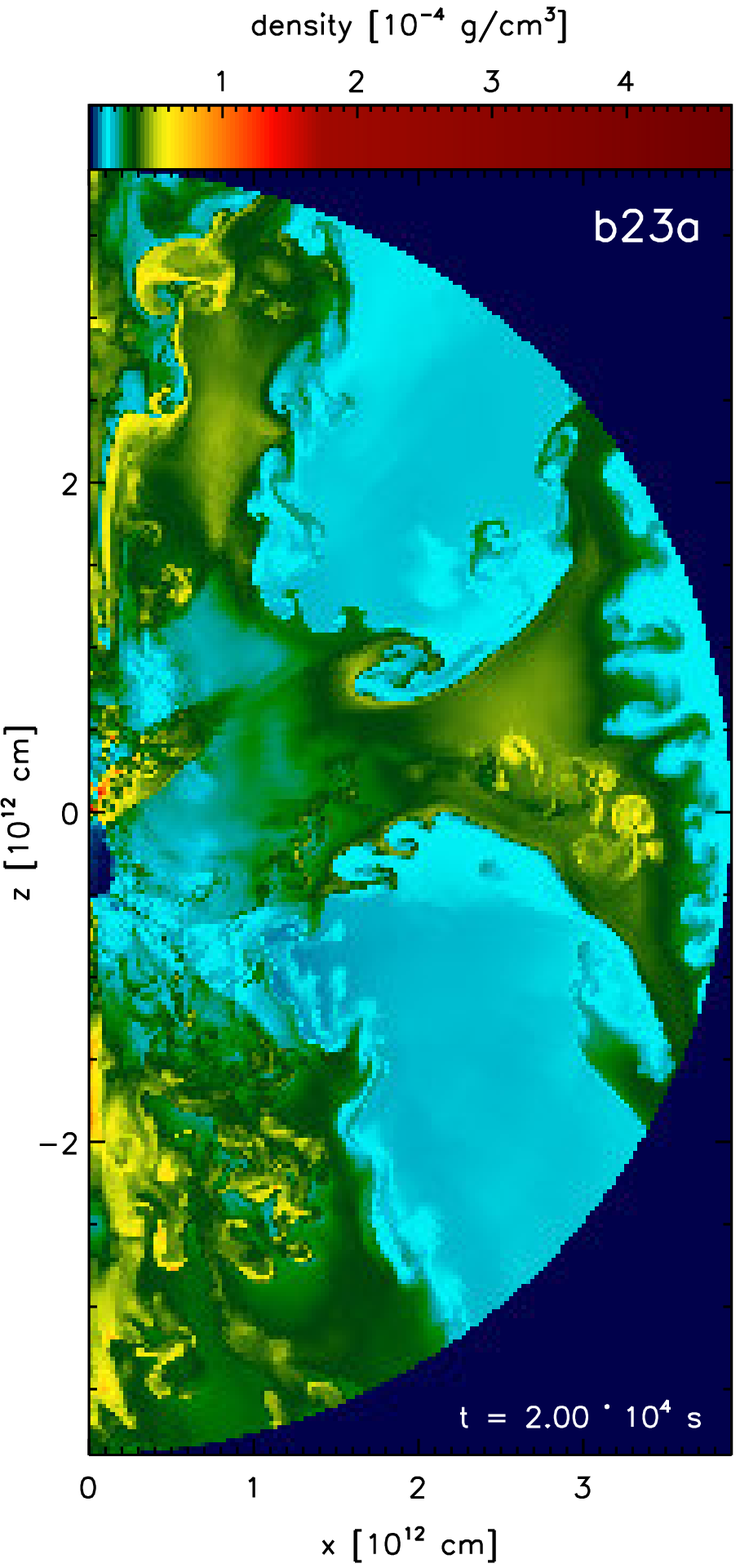} \\
\end{tabular}
\caption{Same as Fig.~\ref{fig:b23a_dens_a}, but for selected late
         instants in time. From left to right a) $t \approx 1500$\,s,
         b) $t \approx 10\,000$\,s, and c) $t \approx 20\,000$\,s.
         The outer circular boundary in the last two plots is the
         outer boundary of the computational domain, which the
         supernova shock has already left at those times.  The shock
         is only visible in the plot for $t \approx 1500$\,s and has
         become almost completely spherical at this time, while it is
         decelerating in the hydrogen envelope. Below it, the
         Richtmyer-Meshkov instability at the He/H interface has
         grown, while in still deeper layers a strong reverse shock
         decelerates the matter of the He core. Note (by comparing
         Fig.~\ref{fig:b23a_dens_b}a with Fig.~\ref{fig:b23a_dens_a}f)
         that a few of the Rayleigh-Taylor clumps have actually
         reached the He/H interface before this reverse shock has
         managed to form, and that thereby they have escaped an
         interaction with this shock. Note also the huge hydrogen
         pockets that are created by the Richtmyer-Meshkov
         instability.}
\label{fig:b23a_dens_b}
\end{figure*}

\begin{figure*}[tpb!]
\centering
\begin{tabular}{ccc}
\includegraphics[width=5.2cm]{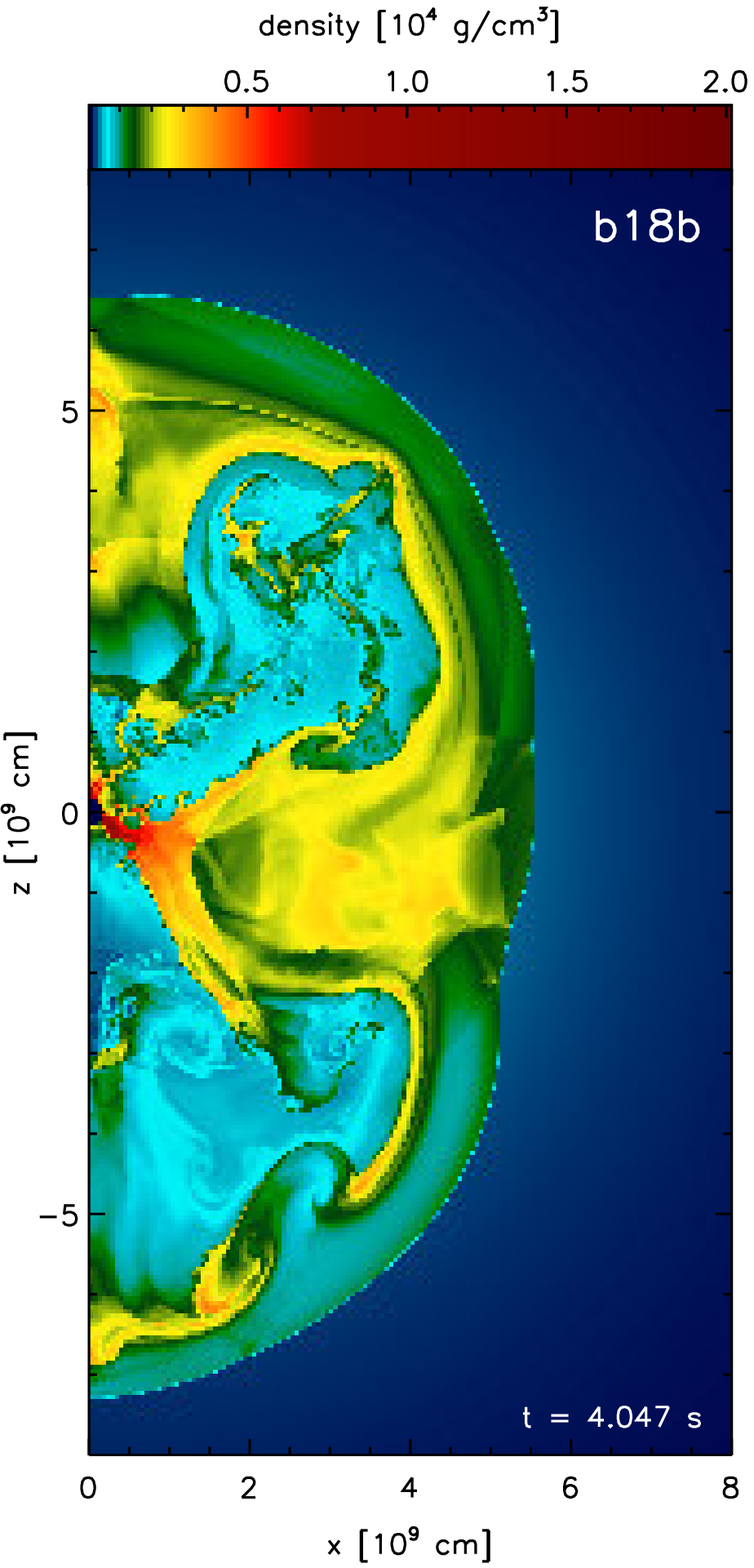}  &
\includegraphics[width=5.2cm]{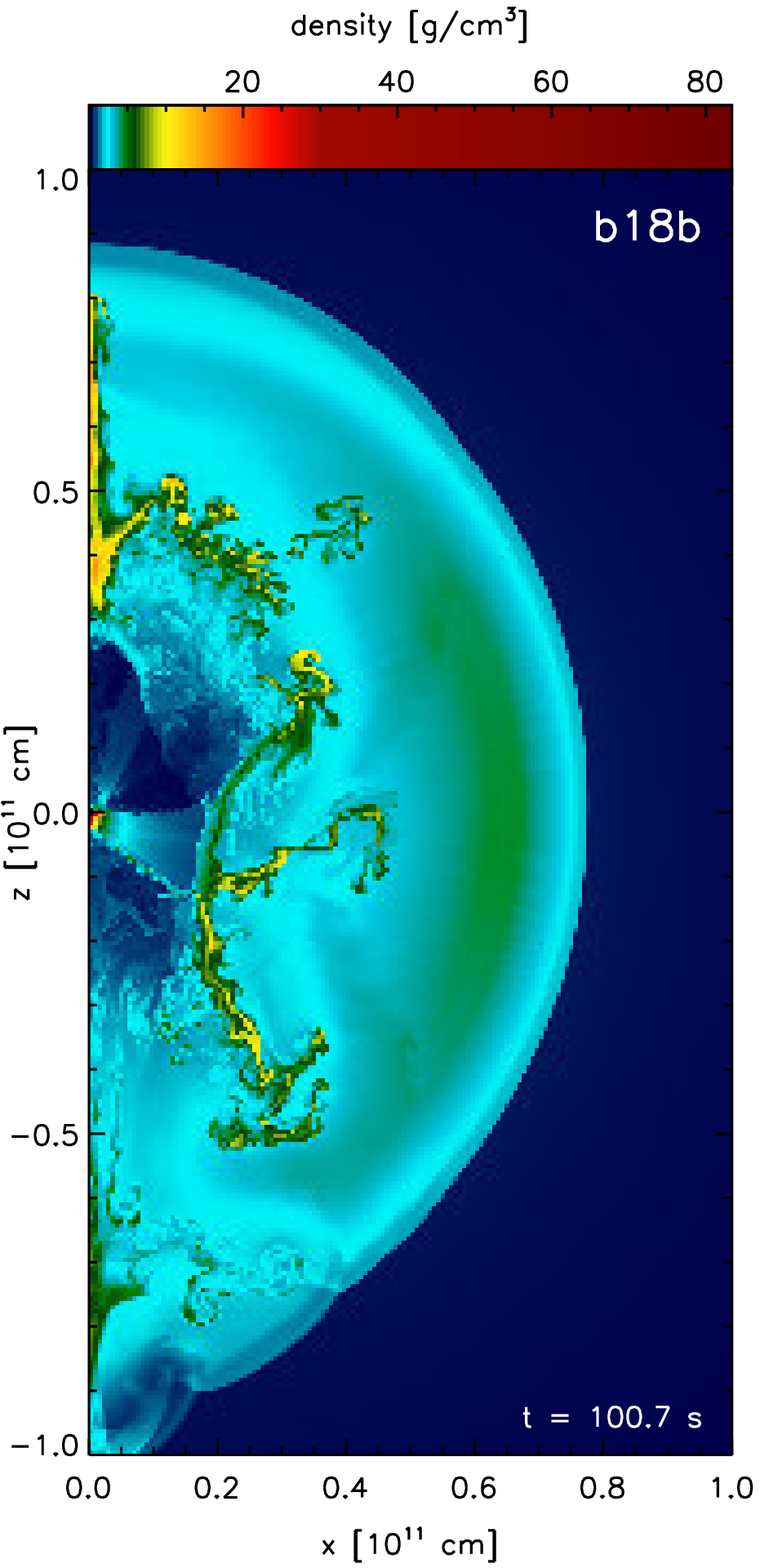}  &
\includegraphics[width=5.2cm]{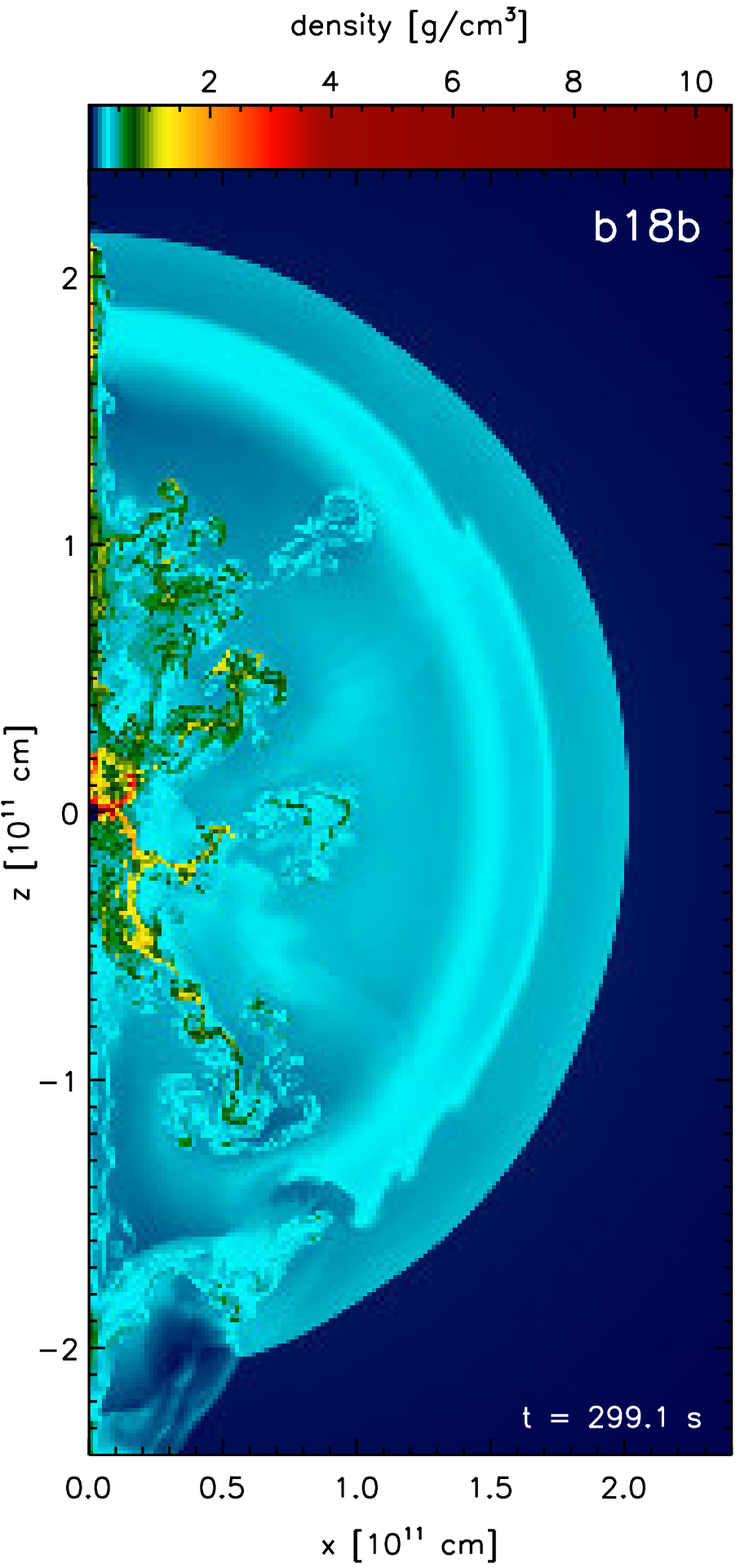}  \\
\includegraphics[width=5.2cm]{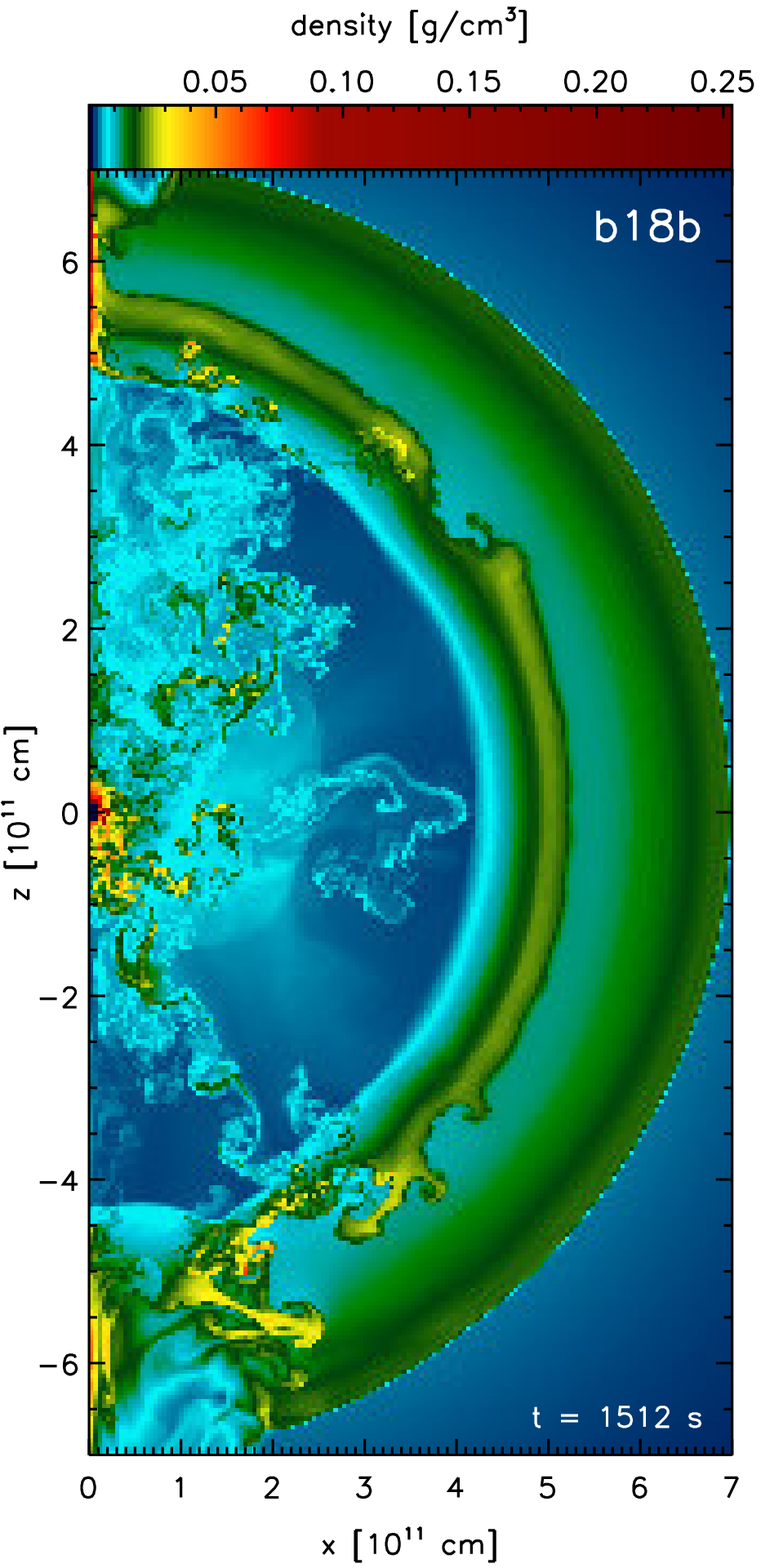}  &
\includegraphics[width=5.2cm]{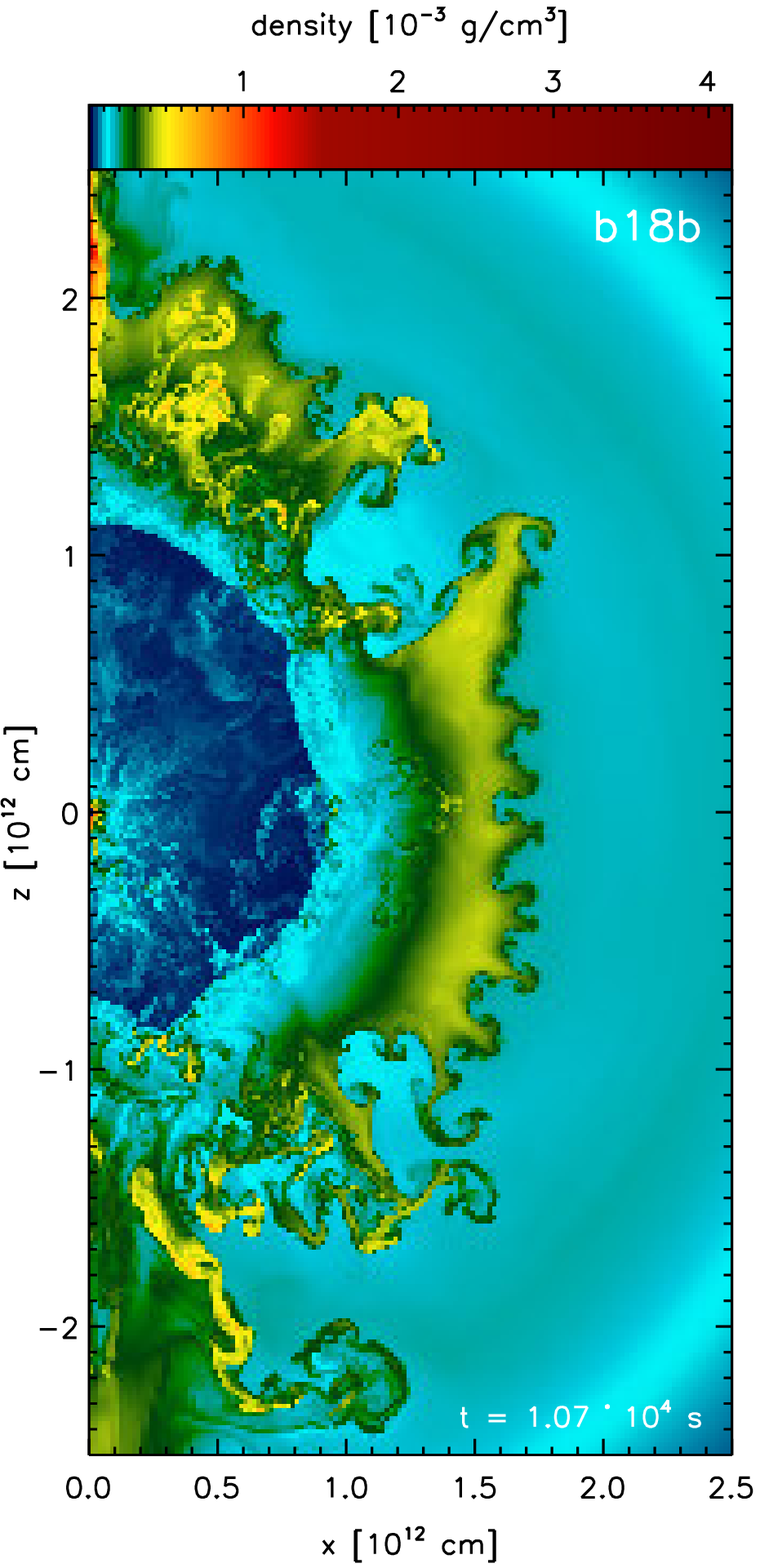}  &
\includegraphics[width=5.2cm]{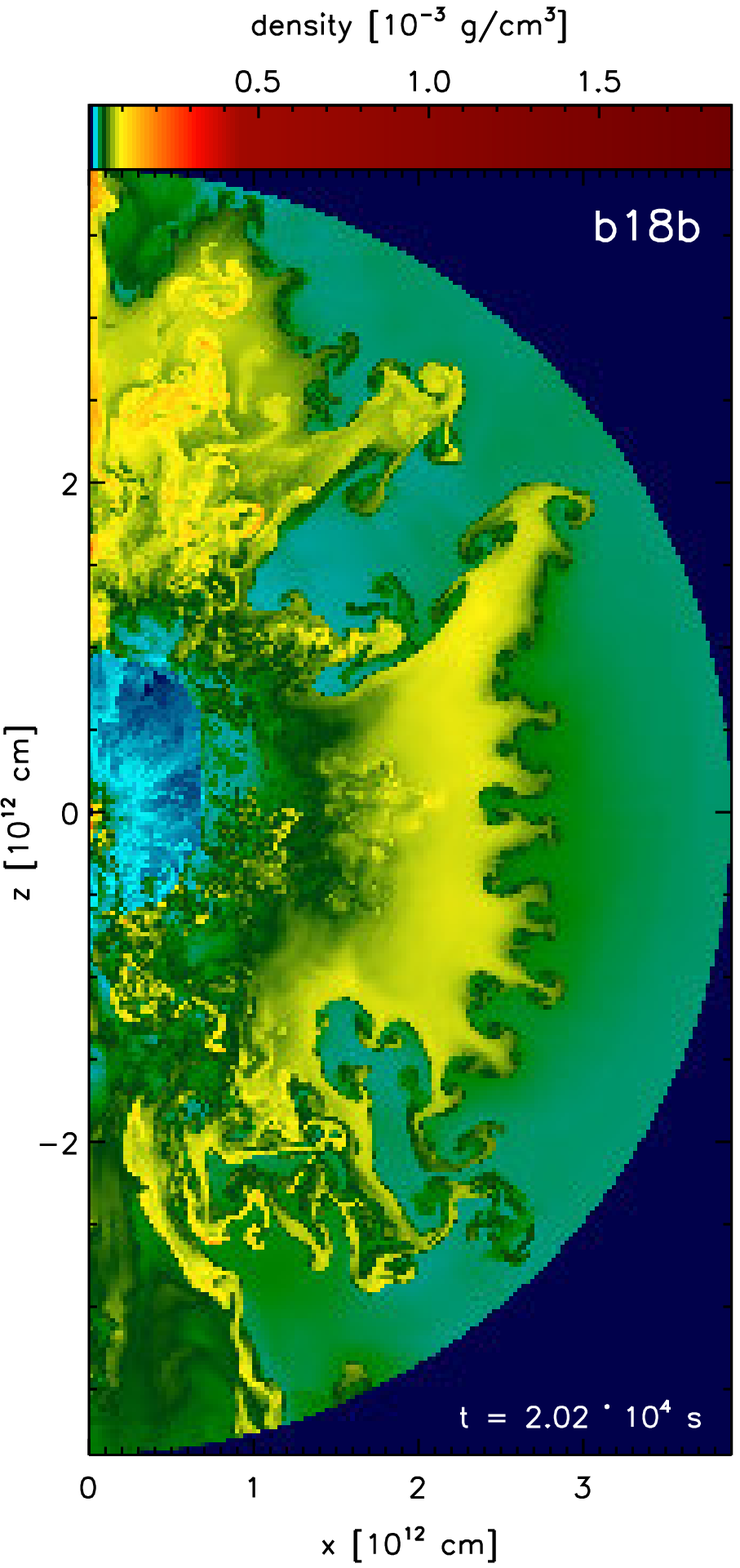}  \\
\end{tabular}
\caption{Snapshots of the density distribution of Model b18b for
         several times. From top left to bottom right a) $t=4$\,s, b)
         $t=100$\,s, c) $t=299$\,s, d) $t=1512$\,s, e) $t \approx
         10\,000$\,s, and f) $t \approx 20\,000$\,s. Note the change
         of the radial scale. The supernova shock is visible in plots
         a)--d) only. The outer circular boundary in plot f) is again
         the outer boundary of the computational domain. Note also the
         more pronounced hemispheric asymmetry of the ejecta at late
         times as compared to Model b23a shown in
         Fig.~\ref{fig:b23a_dens_b}.}
\label{fig:b18b_dens}
\end{figure*}

\subsection{Energetics, timescales, and morphology}
\label{sect:time_scales}

Table~\ref{tab:models_blue} provides an overview of all models that we
have computed for the $15\,\Msol$ blue supergiant progenitor of
\cite{WPE88} to date. Model b18a (which we will not discuss further
due to the reasons mentioned in Sect.~\ref{sect:grid_constraints}) and
Model b18b are ``medium core luminosity'' models with an explosion
energy of $10^{51}$~erg. They differ only in the random seed
perturbation that we added to the velocity field of Bruenn's
post-bounce model to trigger the onset of neutrino-driven
convection. They differ hardly in their global parameters like the
explosion energy, but show large differences in the morphology,
indicating the extremely nonlinear character of the growth of
non-radial instabilities in the SN core. Model b23a, on the other hand,
is a ``high core luminosity'', high-energy explosion of
$2\times10^{51}$~erg. We recall once more that in all of these
simulations the core luminosities were held constant for the first
second after core bounce (see Sect.~\ref{sect:setup}).

In contrast, Model T310a (which has been described in detail in
Paper~I) employed core luminosities that declined exponentially with
time. We will include it in the following discussion in order to
demonstrate that our new calculations show dramatic quantitative and
qualitative differences compared to this older simulation. We strongly
encourage the reader to compare the density and velocity plots of
Model T310a that we presented in Paper~I, with the plots for Model
b23a (Figs.~\ref{fig:b23a_dens_a}, \ref{fig:b23a_dens_b} and
\ref{fig:massvelo_b23a}) and Model b18b (Figs.~\ref{fig:b18b_dens} and
\ref{fig:massvelo_b18b}) that we show in this paper. The different
morphology of the neutrino heated ejecta and the shock within the
first few seconds after core bounce is striking. As already stated in
the introduction, this is basically a consequence of different
explosion timescales and the thus different growth of non-radial
instabilities during the shock revival phase by neutrino heating, and
the onset of the explosion.

With a timescale $t_{\rm exp}$ of only 62\,ms, Model T310a exploded
fast -- nearly as fast as the typical turnover time of a convective
eddy. This neither allowed convection nor the advective-acoustic shock
instability to grow appreciably. Only \emph{small} convective blobs
were formed in the neutrino-heated layer out of the imposed random
perturbations, and the shock wave remained undistorted
(i.e. spherical). Moreover the flow pattern was dominated by a
high-order mode, with about a dozen eddies.

In contrast, Models b23a and b18b exploded more slowly by a factor of
between two and three (Table~\ref{tab:models_blue}). Hydrodynamic
instabilities in the post-shock flow thus had more time to grow. As in
the simulations of \cite{Scheck+04,Scheck+06} the initially small
blobs in the neutrino-heated layers merged, and ultimately formed a
low-order, $l=2$ dominated mode with a smaller $l=1$ mode
contribution, exhibiting only two large buoyant bubbles which are
separated by a single accretion funnel. In the process, they deformed
the shock wave to a prolate, almost peanut-like shape (see
Figs.~\ref{fig:jets}b, \ref{fig:jets}c, \ref{fig:b23a_dens_a}a, and
\ref{fig:b18b_dens}a). 

The longer explosion time scales also led to larger neutron star
masses in Models b23a and b18b around one second after core bounce
than in Model T310a (Table~\ref{tab:models_blue}). While these masses
are still rather low, the comparison shows that the exact value is
fairly sensitive to the treatment of the neutrino physics and the
assumed boundary conditions (i.e. the neutrino properties and the
contraction behaviour of the neutron star core). When the explosion
sets in later, the neutron star mass is higher. Moreover it also
depends on the structure of the progenitor star (see
\citealt{Scheck+06} for the corresponding significant variations in
case of different $15~\Msol$ progenitors), and can increase due to
later fallback. We have not traced this fallback in the present work,
though.

\subsection{Shock expansion and Richtmyer-Meshkov instability}
\label{sect:Richtmyer}

The early global anisotropy of the shock in the low-mode models is
decisive for the further evolution. It leads to qualitatively new
effects in the late-time hydrodynamics compared to Model T310a, in
which the shock remained spherical from the outset, or the simulation
presented in \cite{Kifonidis+00}, where the shock had been deformed by
rising bubbles during the neutrino-heating phase, but had become
spherical by the time it emerged from the He core (see
\citealt{Chevalier_Soker89} for a discussion of the underlying
mechanism). Note that in our previous simulations such deformations of
the shock due to buoyant neutrino-heated bubbles were in all cases
\emph{local} ones, with deviations of the local to the mean shock
radius of at most 20\%. In contrast, Model b23a has a prolate shock
with an initial major to minor axis ratio of $\sim 1.5$.  A
deformation of this magnitude not only makes the definition of a mean
shock radius questionable. It is so substantial that it apparently
cannot be smoothed out completely by the time the blast arrives at the
He/H interface of the \citeauthor{WPE88} progenitor.

In Model b23a the original peanut-like shock gives way to a surface
with an equatorial bulge, that forms at the ``peanuts' waist'' a few
seconds after core bounce, and progressively attempts to bring the
shock into the spherical shape (Fig.~\ref{fig:b23a_dens_a}). Yet, it
requires some time to succeed in this, and when the blast has crossed
the He/H interface around $100$\,s after core bounce its surface still
shows two kinks. Kinks in the shock surface are usually found above
prominent downflows, and our ``two-kink feature'' originated from the
merging of two such downflows.

The kinks in Model b23a are located at similar latitudes in the
northern and southern hemispheres (see
Fig.~\ref{fig:b23a_dens_a}e). In these spots the shock hits the
composition interface obliquely. As a result, it deposits a
significant amount of vorticity into the interface layer, which
triggers strong Richtmyer-Meshkov mixing
\citep{Richtmyer_60,Meshkov_69}, much as in the simple, planar
geometry setup considered by \cite{Hawley_Zabusky89} (in fact, our
Fig.~\ref{fig:b23a_dens_b}a shows all features that are visible in the
plots of their flows, in particular of the ``fast-slow'' setup that
they studied). Two prominent vortices\footnote{In the context of
axisymmetric simulations all flow features are actually toroidal
structures, and the ``vortices'' are actually ``vortex rings''.} form,
which transport hydrogen-rich gas in the form of two large ``pockets''
into the helium and metal core. At the same time material from these
core layers is dragged outward into the hydrogen envelope, while a
vortex sheet due to the Kelvin-Helmholtz instability forms along a
significant part of the interface (Figs.~\ref{fig:b23a_dens_b}a--c).

\begin{figure}
\centering
\includegraphics[width=\linewidth]{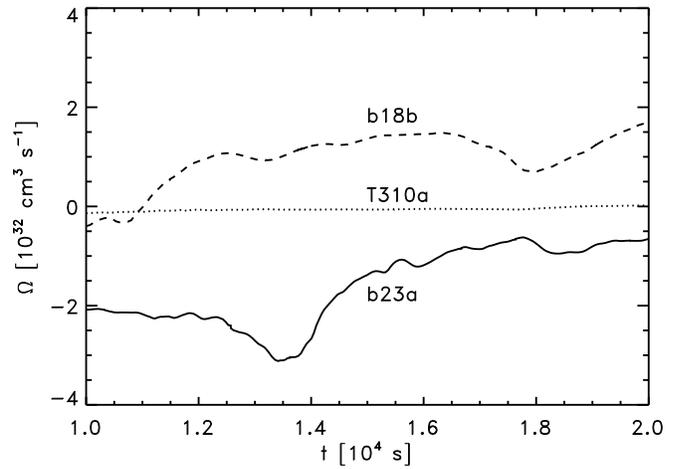} 
\caption{Integrated vorticity $\Omega$ according to
         Eq.~(\ref{eq:omega}) as a function of time during the late
         evolutionary phases of Model T310a (dotted line), Model b23a
         (solid line), and Model b18b (dashed line). Note that
         $\Omega$ is essentially zero for the high-mode model T310a,
         while it is of the order $10^{32}\,{\rm cm^3 s^{-1}}$ for the
         low-mode models b23a and b18b, which experience a strong
         Richtmyer-Meshkov instability at the He/H interface.}
\label{fig:vort}
\end{figure}

Note that the He/H interface is actually susceptible to both, the
Richtmyer-Meshkov, and the Rayleigh-Taylor instability, which is due
to pressure and density gradients of opposite sign in this region (see
Paper~I). At this interface, however, Richtmyer-Meshkov instabilities
grow faster and appear earlier than RT mushrooms\footnote{The
Richtmyer-Meshkov instability is an essentially impulsive instability
triggered upon shock passage. The RT instability is taking place on a
much longer timescale due to the smooth large-scale gradients of
pressure and density at the He/H interface.} (compare
Figs.~\ref{fig:b23a_dens_b}b and \ref{fig:b23a_dens_b}c). On the long
run the mixing in fact is a result of both instabilities interacting
with each other. Yet, we will still refer to the mixing at the He/H
interface (a bit loosely) as ``Richtmyer-Meshkov instability'' in
order to make clear that it is the deposition of vorticity by the
aspherical shock which is crucial. To demonstrate this, we consider
\begin{equation}
\boldsymbol{\Omega } = 
                     \int_{V} \left( \nabla \times \boldsymbol{u}
                              \right) \, {\rm d}V,
\label{eq:circ}
\end{equation}
the integral of the vorticity over the computational domain. In
Eq.~(\ref{eq:circ}) $\boldsymbol{u}$ is the flow velocity, $V$ is the
computational volume, and ${\rm d}V$ is the volume element. For 2D
axisymmetric calculations in spherical coordinates there is only one
component of $\nabla \times \boldsymbol{u}$ that does not vanish,
namely $\left(\nabla \times \boldsymbol{u}\right)_{\varphi}$, and
therefore $\boldsymbol{\Omega}$ reduces to the scalar
\begin{equation}
\Omega = 2 \pi
\int_{R_{\rm i}}^{R_{\rm o}} \negthickspace
\int_{0}^{\pi}
  {1 \over r} \left[ {\partial   \over \partial r} (r v ) -
                              {\partial u \over \partial \vartheta}
                       \right] r^2 \sin{\vartheta} \,
                               {\rm d}r \, {\rm d}\vartheta,
\label{eq:omega}
\end{equation}
where $u$ and $v$ are the velocity components in $r$ and $\vartheta$
direction, and $R_{\rm i}$ and $R_{\rm o}$ are the inner and outer
(open) radial boundaries, respectively. Note that $\Omega$ is a
suitable measure for the strength of vortices residing in the domain.

\begin{figure}
\centering
\begin{tabular}{c}
\includegraphics[width=0.95\linewidth]{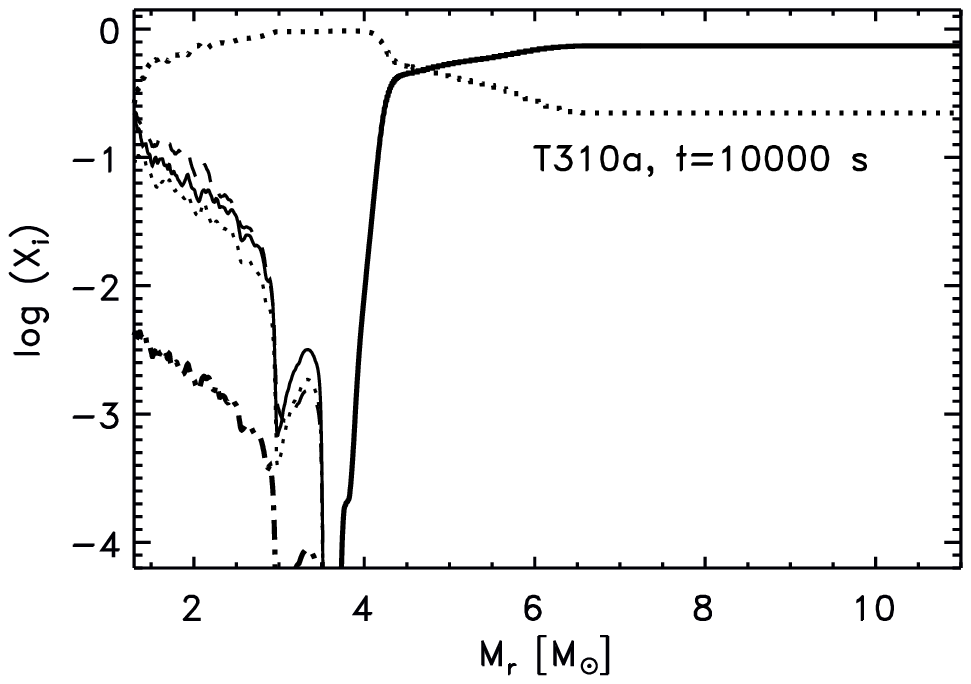} \\
\includegraphics[width=0.95\linewidth]{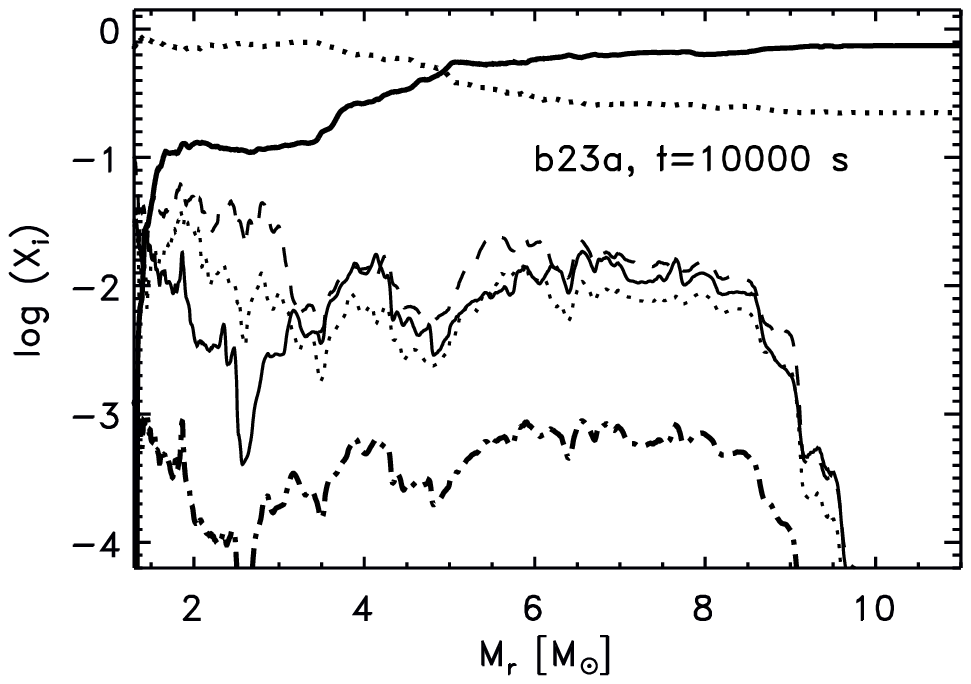} \\
\includegraphics[width=0.95\linewidth]{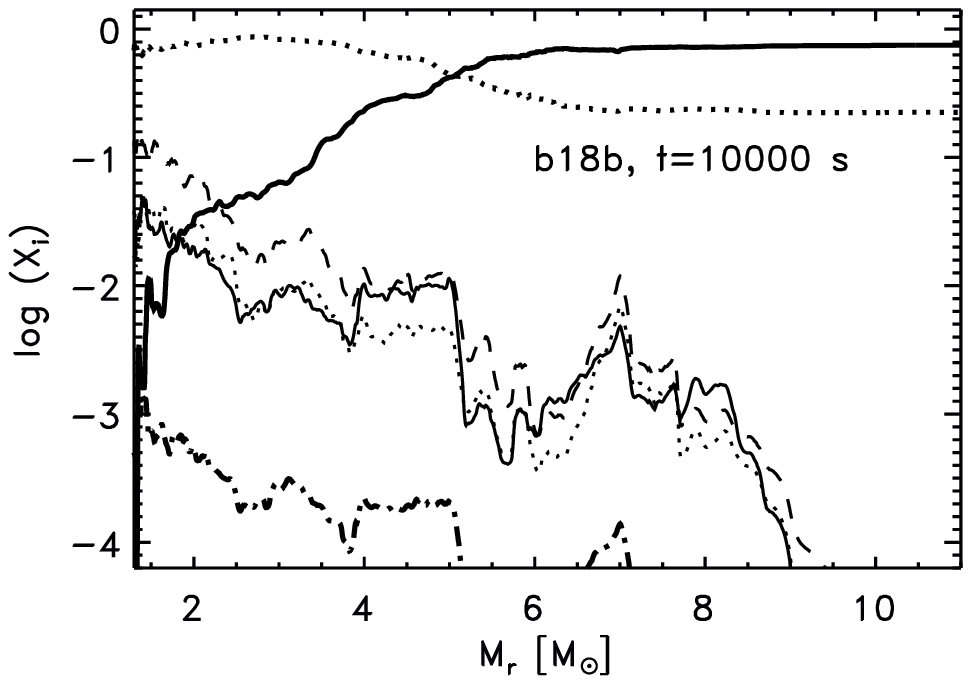} \\
\end{tabular}
\caption{Extent of the chemical mixing in the simulations at a time of
         10\,000\,s. From top to bottom a) Model T310a, b) Model b23a,
         and c) Model b18b. In all cases the angle-averaged mass
         fractions of $\rm H$ (bold solid line), $\rm ^{4}He$ (bold
         dotted line), $\rm ^{16}O$ (thin dashed line), $\rm
         ^{28}Si$ (thin dotted line), $\rm ^{44}Ti$ (bold
         dashed-dotted line), and the iron group (thin solid line), are
         shown as a function of the enclosed mass. Note the deep
         inward mixing of hydrogen and the strong outward mixing of
         the metals in Models b23a and b18b, and the lack thereof in
         Model T310a.}
\label{fig:mixing}
\end{figure}

Figure~\ref{fig:vort} shows the temporal evolution of $\Omega$ for
Models b23a, b18b, and T310a, and times between $10^{4}$ and
$2\times10^{4}$\,s. When the shock has already left the grid (which is
the case at these times), no contamination of $\Omega$ due to the
axial artifacts visible in Fig.~\ref{fig:jets} is present any more,
and the only significant contribution stems from the vorticity
deposited early on by the shock at the He/H interface. This
contribution is of the order of a few $10^{32}\,{\rm cm^3 s^{-1}}$ for
Models b23a and b18b, while it is essentially zero for Model T310a
with its spherical shock.

\begin{figure*}
\centering
\begin{tabular}{c}
\includegraphics[width=0.8\linewidth]{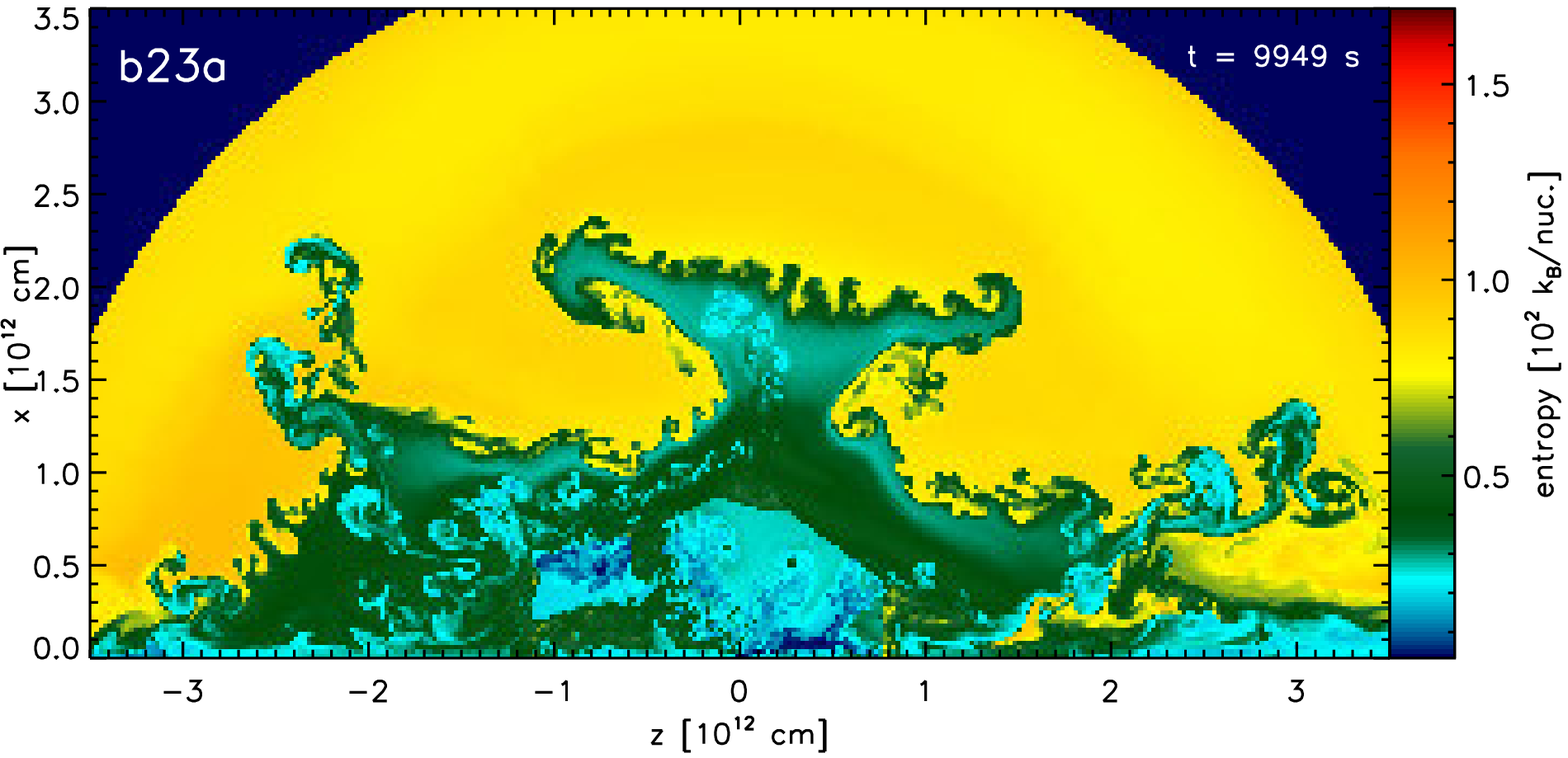} \\
\includegraphics[width=0.8\linewidth]{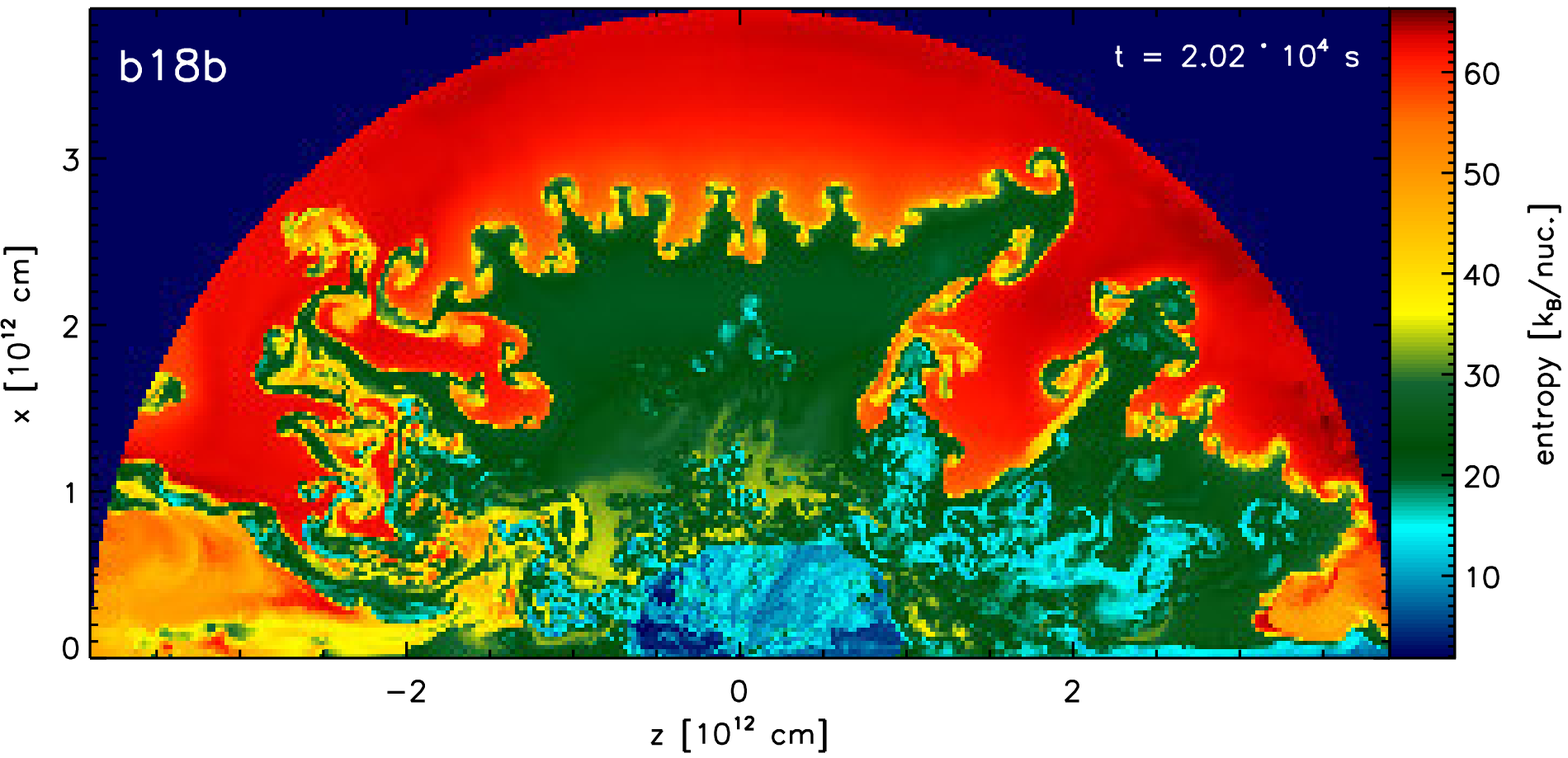} \\
\end{tabular}
\caption{Entropy distribution (in units of $k_{\rm B}$/nucleon) of the
         low-mode models at late times. From top to bottom: a) Model
         b23a at $t \approx 10\,000$\,s and b) Model b18b at $t
         \approx 20\,000$\,s. The color coding has been chosen such
         that material with different composition can be clearly
         distinguished. Hydrogen-rich gas is shown in orange or red,
         while material of the helium core is coded in dark
         green. Matter from the metal core, which includes nuclei that
         have been synthesized during the explosion as well as during
         the hydrostatic evolution of the star, is shown in light and
         dark blue.}
\label{fig:stot}
\end{figure*}

\begin{figure*}
\centering
\includegraphics[width=15cm]{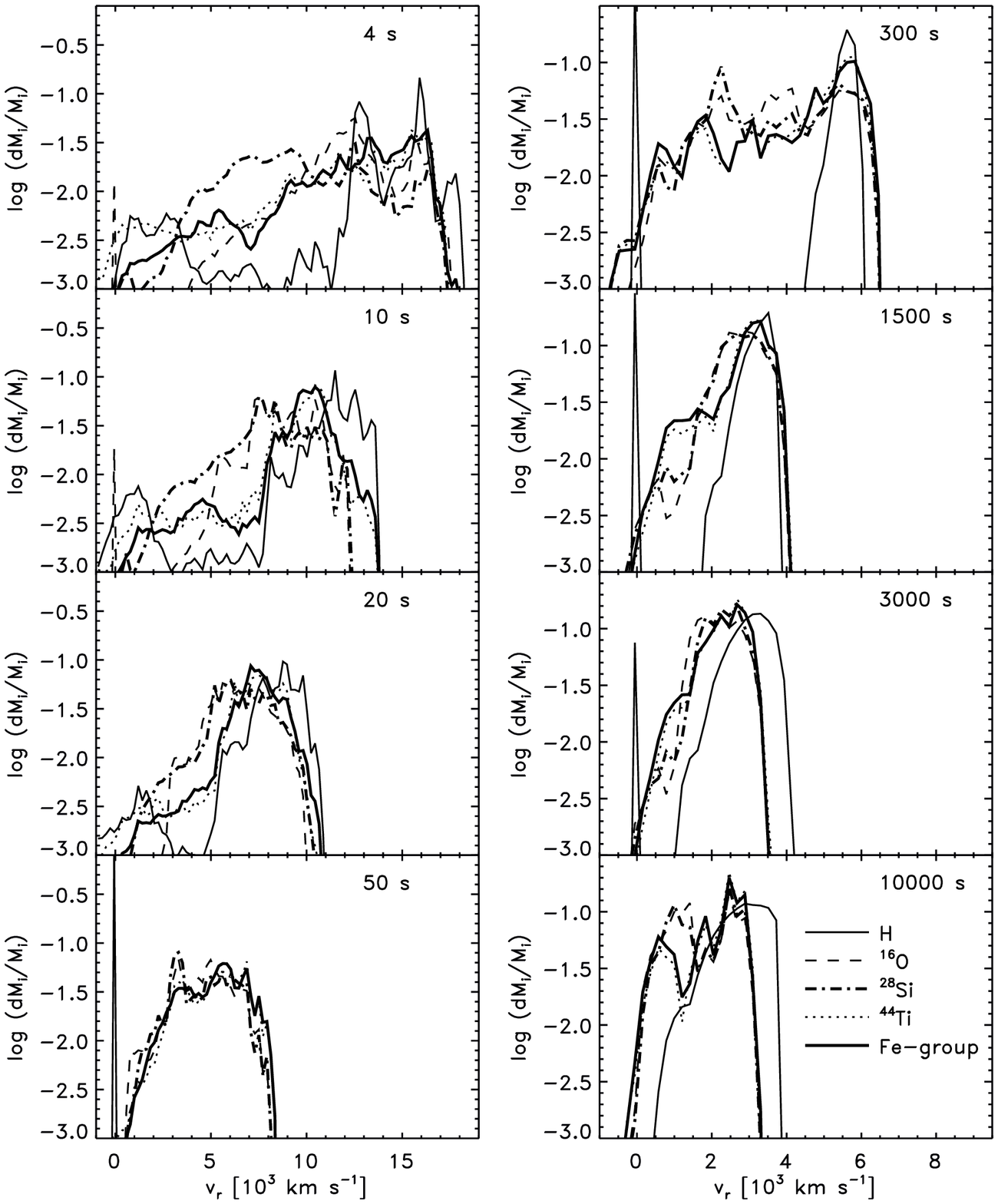}
\caption{Logarithm of the fractional mass of different elements as a
         function of the radial velocity $v_r$ for Model b23a at
         various epochs. Note the different scales for $v_r$ in the
         left and right panels. Note also that the fractional mass was
         computed with respect to the mass of the element, $M_i$, that
         is contained on the \emph{current} extent of the
         computational grid. For the lighter nuclei H and $\rm
         ^{16}O$, this mass is \emph{not necessarily} identical to the
         total element mass, because initially our adaptive mesh does
         not yet cover the outermost stellar layers, and because in
         the late phases the shocked gas is allowed to stream off the
         grid. The latter fact is the reason why the hydrogen velocity
         distribution at $t = 10\,000$\,s does not extend beyond $\sim
         4000$\,km/s (because faster expanding hydrogen has left the
         grid already).}
\label{fig:massvelo_b23a}
\end{figure*}

\begin{figure*}
\centering
\includegraphics[width=15cm]{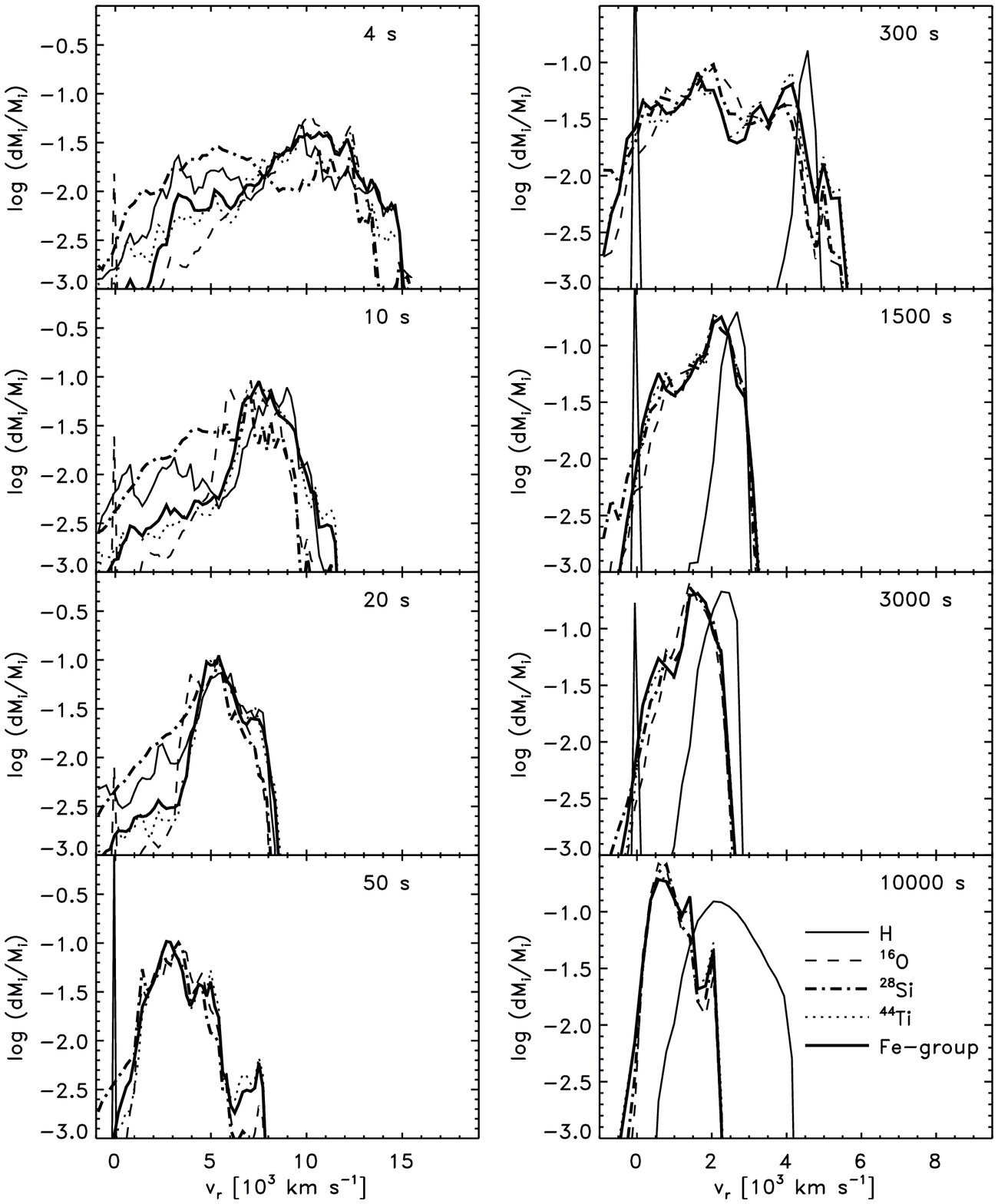}
\caption{Same as Fig.~\ref{fig:massvelo_b23a}, but for Model b18b.}
\label{fig:massvelo_b18b}
\end{figure*}

We have already shown in Paper~I that Model T310a exhibited only a
very weak Rayleigh-Taylor instability at the He/H interface, which was
triggered by perturbations of this interface due to acoustic noise
(i.e. sound waves). Although this led to the growth of a large number
of small Rayleigh-Taylor mushrooms, Model T310a clearly \emph{failed}
to mix significant amounts of hydrogen into the helium and metal core,
i.e. below a mass coordinate of $\sim 4\,\Msol$ (compare
Fig.~\ref{fig:mixing}a). The ``high-vorticity'' models, b23a and b18b,
on the other hand, succeed in mixing hydrogen down to a mass
coordinate of $\sim 1.3\,\Msol$ (Figs.~\ref{fig:mixing}b and c), which
is close to the neutron star mass. This clearly demonstrates that in
addition to the short-wavelength Rayleigh-Taylor instability (which
can be easily identified in the late evolutionary stages of Models
b23a and b18b by the typical mushroom-shaped structures, see
Figs.~\ref{fig:b23a_dens_b}, \ref{fig:b18b_dens} and \ref{fig:stot}),
a much more efficient vortical mixing mechanism is present in these
models, which operates on much larger scales. Here we identify this
vortical mechanism with the Richtmyer-Meshkov instability \citep[see
also][and the references therein]{Hawley_Zabusky89}.

A consequence of the nature of this instability is that it leads to a
dependence of the final (large-scale) spatial distribution of chemical
elements on the initial shape of the shock, i.e. on the spectrum of
the superposed unstable modes in the early post-shock flow
pattern. If, initially, two equally large buoyant high-entropy bubbles
are present behind the shock, which are separated by a single
prominent accretion funnel that is located close to the equatorial
plane, i.e. if a dominant $l=2$ mode is present (see
Figs.~\ref{fig:jets}c and \ref{fig:b23a_dens_a}a), then the later
distortion of the shock surface still reflects this initial asymmetry
of quadrupolar character.  In this case two nearly equally large
pockets of hydrogen finally develop, which, together with the
distribution of the outward dragged material from the helium and metal
core, show almost equatorial symmetry (Fig.~\ref{fig:stot}a). If on
the other hand, the buoyant bubbles are of different size and the
accretion funnel is bent and inclined towards one of the poles,
i.e. if there is some admixture of the $l=1$ mode in the initial flow
pattern, as in Model b18b (see Figs.~\ref{fig:jets}b and
\ref{fig:b18b_dens}a), the H pockets tend to grow to different
sizes. The final hydrogen distribution then develops a pronounced
hemispheric asymmetry, as do also the spatial distributions of helium
and the metals (Fig.~\ref{fig:stot}b).

Both the deep inward mixing of hydrogen, and the natural development
of large-scale, hemispheric asymmetries in the spatial distribution of
different elements, have important consequences for the interpretation
of observational data of supernova explosions and supernova
remnants. The former is required to obtain good fits in light curve
modelling, while the latter is probably the cause of the asymmetric
iron (and nickel) lines of SN 1987\,A. We will return to these issues
in Sect.~\ref{sect:SN_1987A}.

It is quite astonishing that (unlike the RT instability) the
importance of the Richtmyer-Meshkov instability for the observational
appearance of core collapse supernova explosions has not been widely
recognized in previous modelling (but see e.g. \citealt{Kane+00} for an
exception). This may be, at least in part, due to the fact that in
previous studies of aspherical shock propagation through exploding
stars this instability was not noticed. Judging from the published
information, it is for instance not visible in the calculations of
\cite{Hungerford+03,Hungerford+05}. These authors started their 3D
simulations at a time of 100\,s after core bounce from a 1D
calculation of the earlier phases of the explosion. Thus the shock had
already crossed the He/H interface at the time they added asymmetries
to the velocity field of their spherically symmetric explosion
models. This obviously gave the instability no chance to develop.

In the works of \cite{Nagataki00} and \cite{Yamada_Sato91}, on the
other hand, who initiated their 2D simulations from aspherical shocks
at much earlier times, it was probably the coarse angular resolution
of only 100 zones which prevented them from discovering the
instability (in the latter paper a diffusive advection scheme with
only first-order accuracy in space was employed, too). We think
that this demonstrates the importance of two points which we have
already addressed in Paper~I, namely that modelling the mixing in
core-collapse SNe requires one to follow the evolution from the
\emph{earliest} phases of the explosion in more than one dimension (in
order to avoid biasing the late-time evolution). Furthermore, high
resolution and an excellent advection scheme are essential for
resolving and properly describing the relevant fluid instabilities.

\subsection{Metal clumps and Rayleigh-Taylor instability}

The unstable modes that become dominant within the first second of the
explosion, do not only determine the morphology of the shock, and
thereby the sequence of events at the He/H interface. They also
provide the perturbations for the Rayleigh-Taylor unstable Si/O and
O/He interfaces of the progenitor. We have noticed in Paper~I that the
wavelength of these perturbations determines the number of
Rayleigh-Taylor fingers that grow at these interfaces, and that it
characterizes the flow pattern until late times. This is also supported
by the recent work of \cite{Miles04}, who has developed a
Rayleigh-Taylor bubble merger model for compressible flows. He
concluded that unless the shock Mach number and the mode number of the
initial perturbation are very high, the system will not loose memory
of the original perturbation spectrum.

This is indeed the case in all models that we have computed to date.
Model T310a developed about a dozen Rayleigh-Taylor fingers out of a
comparable number of flow features that formed during the phase of
neutrino-driven convection near the Si/O and O/He interfaces
(Paper~I). The low-mode models b23a and b18b, on the other hand, lead
to the growth of only about three to four, rather large,
Rayleigh-Taylor mushrooms (Figs.~\ref{fig:b23a_dens_a} and
\ref{fig:b18b_dens}). In what follows we will keep referring to the
case of Model T310a as a high-order, and to this of Models b23a and
b18b as a low-order mode, in order to distinguish them clearly. In the
sense of \cite{Miles04} all of our models bear ``low-mode''
perturbations, because they all satisfy his condition for retainment
of the initial mode number.

Besides having a different number and size of Rayleigh-Taylor
``clumps'', both cases also differ strongly in the velocity
distributions of nucleosynthetic products. The low-mode models show
much broader distributions in velocity space for iron-group nuclei and
for the isotopes $\rm ^{16}O$, $\rm ^{28}Si$ and $\rm ^{44}Ti$ a few
seconds after bounce than Model T310a (compare the plots for $t = 4$~s
in Figs.~\ref{fig:massvelo_b23a} and \ref{fig:massvelo_b18b} with
Fig.~16 of Paper~I).

Even more importantly, the \emph{maximum velocities} of
these nuclei are significantly larger than in the high-mode model
T310a. In Model b23a we find maximum metal velocities 4 seconds after
bounce that are about 40\% larger than in Model T310a. But even in
the low-energy, low-mode model b18b, whose explosion energy is nearly
half that of Model T310a ($1.0$ vs. $1.7\times 10^{51}\,{\rm erg}$,
respectively), the corresponding velocities are by about 20\% larger.

The spatial distribution of kinetic energy in the low and high-mode
cases is fundamentally different. In Model T310a most of the kinetic
energy was contained in an essentially spherical shell behind the
shock (similar to a 1D simulation), and there was only little kinetic
energy in the two-dimensionally perturbed region containing the
bubbles. The weak neutrino-driven convection in Model T310a was not
able to boost the velocities inside those bubbles relative to the mean
expansion velocity of the background flow. In the low-mode models this
is different. The high-entropy bubbles and the higher-density pockets
enclosed by them contain most of the kinetic energy of the explosion.

The higher metal velocities during the first few seconds after core
bounce lead to significant differences in the late-time evolution of
the low-mode models as compared to Model T310a. The metal clumps now
propagate faster through the He core of the star, keeping closer
contact to the outward sweeping supernova shock. In both, Model b23a
and Model b18b, the fastest clumps have almost reached the He/H
interface around 300\,s after core bounce (see
Figs.~\ref{fig:b23a_dens_a}f and \ref{fig:b18b_dens}c). At this time
the strong reverse shock that will develop below this interface due to
the deceleration of the main shock in the hydrogen envelope (compare
Fig.~\ref{fig:b23a_dens_b}a and Paper~I), has \emph{not} yet 
formed. Hence, an interaction of the fastest clumps with this reverse
shock, which leads to the dramatic slow-down of the metals beyond a
time of 1500\,s in Model T310a (Paper~I), \emph{does not happen} in
the low-mode models!

Instead, the fastest clumps encounter a layer in which the density
gradient, though steepening, is still relatively flat. Their motion
relative to the background flow thus never becomes supersonic, and
deceleration enhanced by supersonic drag is absent. Hence the maximum
velocities of iron-group nuclei and other elements decrease only
slightly at late times, e.g. from 3500 km/s at $t=1500$~s to 3300 km/s
at $t = 10\,000$~s in Model b23a (Fig.~\ref{fig:massvelo_b23a}). These
velocities remain unchanged until the time when the clumps start to
leave the outer boundary of our grid, which happens at $t \approx
12\,000$~s in Model b23a.

Strong reduction of a late-time deceleration may also be favoured by
the fact that the metal clumps become engulfed in the
Richtmyer-Meshkov instability at the He/H interface. The vortical
energy deposited early on in these layers by the shock, represents a
reservoir of kinetic energy that can possibly be tapped during the
late-time hydrodynamic evolution, and would have been unavailable to
the clumps if the shock had been spherically symmetric when it crossed
the interface (as it was the case in Model T310a).

Summarizing these findings, we can state that by following the
late-time hydrodynamic evolution of neutrino-driven explosions which
were dominated by early, low-mode hydrodynamic instabilities, we have
for the first time obtained a consistent (in the sense defined in
Sect.~\ref{sect:introduction}) solution to the nagging ``nickel
discrepancy'' problem \citep{Herant_Benz92}, i.e. we have obtained
maximum velocities for the bulk of the metal-rich material (in Model
b23a) which are in good agreement with those of SN 1987\,A. This
result calls for a more careful inspection of the simulations in order
to judge whether they also fulfill other observational constraints
obtained from this supernova.

\section{Comparison with SN 1987\,A}
\label{sect:SN_1987A}

The most satisfactory approach for verifying whether our simulations
are consistent with the observational data of SN 1987\,A would, of
course, be the computation of light curves, spectra and the degree of
polarization from the 2D hydrodynamic models. This is beyond the scope
of the present paper and will be attempted in future work.

Here we will confine ourselves to a semi-quantitative comparison with a
check list that we assembled from the results of previous
observational and (one-dimensional) modelling work. This list, which
any successful model of SN 1987\,A needs to pass, consists of the
following items:
\begin{enumerate}
\item
The high velocities of iron-group nuclei, or, given that ultimately
homologous expansion establishes, the outward mixing of these nuclei
in mass. We will focus here on the mixing in velocity space since this
could be measured directly by observations of the Fe and Ni infrared
lines. The mixing in the mass coordinate can be deduced only
indirectly, on the basis of theoretical models, and may thus be less
reliable.

\cite{Haas+90} and \cite{Colgan+94} have carried out observations of
the [Fe~II] $18~\mu{\rm m}$, [Fe~II] $26~\mu{\rm m}$, and [Ni~I]
$6.6~\mu{\rm m}$ lines at 410 and 640 days after the explosion. All of
these lines show broad wings. In addition, there is an isolated
high-velocity emission feature in the red wing of both the iron lines
at day 410, and the nickel line at day 640, which indicates that a
single clump with $\sim3$\% of the total iron mass moves away from the
observer with a velocity of 3900~km/s
\citep{Haas+90}. \cite{Jennings+93} give an expansion velocity of
3150~km/s for this high-velocity feature, based on independent
observations of the [Fe~II] $18~\mu{\rm m}$ line at day 613. The
[Ni~I] $6.6~\mu{\rm m}$ measurements of \cite{Colgan+94} at day 640
exhibited probably the highest signal-to-noise ratio but were affected
by absorption due to dust, which had formed in the ejecta between days
410 and 640. \cite{Colgan+94} thus regard the [Fe~II] lines at day 410
as providing the best evidence for the amount of iron-group elements
at high velocities. From the width of these lines \cite{Haas+90}
concluded that $\gtrsim 4\%$ of the total iron (excluding the
high-velocity feature) has an expansion velocity $\gtrsim 3000$ km/s.

Early nickel mixing with significantly higher velocities (above
5000\,km/s) has been postulated by \cite{Mitchell+01} in order to get
sufficient excitation for the hydrogen Balmer lines in their steady
state spectral models for day 4. \cite{Utrobin_Chugai05}, however,
have recently shown that such high (bulk) nickel velocities are not
required to explain the observations during the photospheric phase, if
one takes into account the effects of time-dependent ionization in
modelling the spectra. With their time-dependent ionization code
excellent fits to the H$\alpha$ profiles are obtained if the $\rm
^{56}Ni$ is mixed out to only $\sim 3000$\,km/s. Further work which
provides compelling evidence of such more moderate velocities for the
bulk of the $\rm ^{56}Ni$ (and, for instance, includes modelling of
the X-ray and $\gamma$-ray emission of SN 1987\,A) is summarized in
the review of \cite{McCray93}. Some more recent papers are the light
curve study of \cite{Utrobin04}, and the work of
\cite{Fassia_Meikle99}. The latter authors have applied spectral
synthesis models to the He~I 10830 \AA\ line after day 10 and found
that about 3\% of the total $\rm ^{56}Ni$ mass must be mixed to above
3000\,km/s, but that the $\rm ^{56}Ni$ concentration must be
negligible above 3900\,km/s. They also demonstrated that models with
this amount of mixing are able to reproduce the $\gamma$-ray light
curves and the late time infrared line velocities.
\item
The marked asymmetry of the [Fe~II] and [Ni~I] infrared lines
\citep{Haas+90,Spyromilio+90,Jennings+93,Colgan+94}. Besides the
(localized) ``high-velocity-feature'' that we already mentioned above,
these lines also show a global red/blue asymmetry. Its origin was
probably anisotropic expansion of the ejecta on a global scale
\citep{Haas+90,Jennings+93}.
\item
The total $\rm ^{56}Ni$ mass of $\sim 0.07\,\Msol$, which is required
to power the light curve
\citep{Woosley88,Suntzeff_Bouchet90,Bouchet+91}.
\item
The very low minimum hydrogen velocities. The mere fact that the
H$\alpha$ line profile observed by \cite{Phillips+90} on day 498 is
not flat-topped implies that there is no large hydrogen cavity in the
slowly expanding central layers of the ejecta (V. Utrobin, private
communication). From the narrow core of the H$\alpha$ line profiles of
\cite{Phillips+90} around day 800, \cite{Kozma_Fransson98} inferred a
minimum hydrogen velocity of $\lesssim 700$~km/s. Mixing of hydrogen
into the SN core, combined with the outward mixing of $\rm ^{56}Ni$, is
also necessary to fit the broad peak of the light curve
\citep[e.g.][]{Woosley88,Shigeyama_Nomoto90,Blinnikov99}. In the
models of \cite{Blinnikov99} hydrogen is mixed down to a mass
coordinate of $\sim 1.3\Msol$.
\item
The prolate deformation of the (inner) ejecta with an axis ratio of
roughly 2:1 that has been found by high resolution imaging with the
HST, and which has been observed indirectly by earlier
spectropolarimetry measurements (see \citealt{Wang+02}, and the
references therein). It is particularly noteworthy that the degree of
polarization was only 0.1\% two days after the explosion and increased
to 1.1\% at about day $200$, indicating that the scattering surface
became increasingly asymmetric as the observations probed the deeper
layers of the ejecta \citep{Wang+02}.
\item
The constraints on the ratio of the explosion energy to ejected mass,
$\eexp/M$, that have been deduced from modelling of the light curve
and the hydrogen lines. \cite{Utrobin_Chugai05} give a value of
$0.83 \times 10^{50}~{\rm erg}~\Msol^{-1}$. \cite{Blinnikov99} finds,
from a multigroup radiation hydrodynamic study, $\eexp/M = (0.75 \pm
0.17) \times 10^{50}~{\rm erg}~\Msol^{-1}$ (based on $\eexp = (1.10
\pm 0.25) \times 10^{51}$~erg, and an assumed ejected mass of
$14.7\,\Msol$), while from the models listed in Table~2 of
\cite{Shigeyama_Nomoto90} an average value of about $0.76 \times
10^{50}~{\rm erg}~\Msol^{-1}$ can be inferred.

\end{enumerate}

We have already noted that in the high-energy model, b23a, the (final)
maximum iron-group velocities are about 3300~km/s. Furthermore, a few
per cent of the total iron-group mass in this model is expanding with
velocities $\gtrsim 3000$\,km/s (compare
Fig.~\ref{fig:massvelo_b23a}), as required by the observations of
\cite{Haas+90}. Our simulations are thus able to reproduce the maximum
velocities of the bulk of the iron-group mass in SN 1987\,A. The
explanation of single, even faster clumps \citep{Utrobin+95} might
require different initial conditions, or higher resolution, or will
have to await future 3D simulations.  Note also that in our low-energy
model, b18b, the highest iron-group velocities are $\sim 2200$\,km/s,
which is too low and indicates that the explosion energy was higher
than $10^{51}$ erg (see also the comments with respect to item six
below).

To be consistent with the second item of the above list, the
simulations are required to exhibit variations of the iron velocities,
and/or the iron concentration with respect to the latitudinal angle on
a large scale, preferably in the form of a hemispheric asymmetry. In
this case one would measure higher velocities, and/or a larger
line-flux from one hemisphere of the ejecta than from the other. Our
models indeed develop such hemispheric asphericities, if a
sufficiently large contribution of the $l=1$ mode to the shock
morphology is present after the neutrino-heating phase. This, on the
other hand, depends on the imposed initial perturbations, which affect
the formation and merging of the neutrino-heated bubbles during the
first $\sim 300$\,ms of the explosion. Since this merging is a chaotic
and highly nonlinear process \citep[cf.][]{Scheck+06}, the development
of a hemispheric asymmetry is in the end governed by
stochastics. Model b18b shows such a pronounced asymmetry, while in
Model b23a (which fits most of the other check-list items better than
Model b18b) such an asymmetry is there, too, but weaker
(Fig.~\ref{fig:stot}). Yet, the outcome might have well been reversed
if different initial perturbations were used. We refrain from
demonstrating this here, as this would require us to calculate a large
sample of models.

A reproduction of item three hinges very sensitively on the electron
fraction, $\ye$, of the neutrino-heated ejecta. The local $\ye$ is
in turn a result of the competition of the charged-current reactions
\begin{eqnarray}
\nue    + n & \leftrightharpoons & e^{-} + p\ , \\
\nuebar + p & \leftrightharpoons & e^{+} + n\ .
\end{eqnarray}
Electron capture on protons and $\nuebar$ absorption by protons
decrease $\ye$ and make the matter more neutron rich. On the other
hand, the inverse processes of positron capture on neutrons and $\nue$
absorption by neutrons drive the matter to the proton-rich side.  The
rates for the $\nue$ and $\nuebar$ captures depend sensitively on the
local $\nue$ and $\nuebar$ spectra, and hence they require an
energy-dependent transport treatment for their accurate
calculation. With a gray transport scheme like ours this can be done
only approximately, resulting in an uncertainty in $\ye$ during
nuclear freezeout of the order of several per cent. Since a variation
of only a few per cent in $\ye$ can result in a more than 100\% change
in the ratio of the $\rm ^{56}Ni$ yield relative to the yields of
other (e.g. more neutron rich) iron-group nuclei, it is not possible
from the present calculations to determine reliable isotopic yield
ratios within the iron-group. We can, however, give an upper limit for
the $\rm ^{56}Ni$ yield, obtained from the $\ye$-insensitive
\emph{total} yield of iron-group nuclei. This upper limit amounts to
$0.15\,\Msol$ and $0.11\,\Msol$ in Models b23a and b18b, respectively,
and is compatible with the fact that $0.07\,\Msol$ of $\rm ^{56}Ni$
were produced in SN 1987\,A.

The consistency of the simulations with item four is easily verified
from Figs.~\ref{fig:massvelo_b23a} and ~\ref{fig:massvelo_b18b}, which
show that the lowest hydrogen velocities in both Models b23a and b18b
are $\sim 500$\,km/s. That the models are also consistent with the
extent of hydrogen mixing assumed in lightcurve modelling is best
checked by comparing our Fig.~\ref{fig:mixing} with Fig.~2 of
\cite{Blinnikov99}. The hydrogen mass fraction profile used by
\citeauthor{Blinnikov99} to obtain a good fit to the observed lightcurve
is almost indistinguishable from the ones that we obtain for
Models b23a and b18b. The sharp drop interior to $\sim 1.5\,\Msol$ and
even the plateau at $\log X = -1$ for mass coordinates $1.5\,\Msol <
M_r < 3.5\,\Msol$ are reproduced excellently.

Consistency with item five can be inspected qualitatively from
Figs.~\ref{fig:b23a_dens_b}, \ref{fig:b18b_dens}, and \ref{fig:stot},
which demonstrate that at late times the inner ejecta (i.e. the helium
and metal-rich material) in our simulations exhibit a global
anisotropy. The major to minor axis ratio of the isodensity contour
corresponding to the He/H interface is roughly 1.6. The decrease of
the asymmetry of the ejecta with radius inferred from the
spectropolarimetry data might be explained from the fact that the
shock in our models (baring the numerical artifacts near the poles)
becomes spherically symmetric once it has propagated through the inner
hydrogen envelope (see the plots for 1500\,s in
Figs.~\ref{fig:b23a_dens_b} and ~\ref{fig:b18b_dens}).  This is an
important feature of our models which clearly distinguishes them from
earlier simulations, e.g. from the asymmetric models presented in
\cite{Yamada_Sato91}. A more detailed comparison with the
spectropolarimetry data -- beyond the qualitative level that we
attempt here -- is difficult because it requires a calculation of the
(time-dependent) location of the photosphere in our models, which we
did not perform.

This leaves us with item six. For our $15\,\Msol$ progenitor and Model
b23a, with its explosion energy of $2.0 \times 10^{51}$ erg, we obtain
an energy-to-mass ratio of $1.4\times 10^{50}~{\rm erg}~\Msol^{-1}$
(after accounting for a neutron star of $1.2\,\Msol$ and ignoring
possible fallback). This is a bit on the high side. For the low-energy
model b18b we get $0.7\times 10^{50}~{\rm erg}~\Msol^{-1}$, which is
in much better agreement with the lightcurve modelling, but for this
model the maximum iron-group velocities are only $\sim 2200$\,km/s,
which is in conflict with item one. Still, we think that with a more
careful choice of the progenitor model and of the explosion energy, it
will be possible to obtain both, agreement with the observed
maximum iron-group velocities, and a closer reproduction of the
$\eexp/M$ ratio because
\begin{itemize}
\item
the mass of the SN 1987\,A progenitor may have been in the
$18-20\,\Msol$ range \citep{Woosley88},
\item
a lowering of the explosion energy from $2.0\times 10^{51}$ erg to
$1.5\times 10^{51}$ erg will lower our maximum metal velocities only
by a factor of $\sim 1.15$ (because these velocities scale
approximately with the square root of the explosion energy),
\item
in three dimensions it is expected that, for the same explosion
energy, larger maximum metal velocities will be obtained (due to the
smaller drag experienced by truly three-dimensional clumps, see
\citealt{Kane+00}).
\end{itemize}

In summary, the low-mode models, and in particular the high-energy
case b23a, are in at least qualitative agreement with the items listed
above. In many cases even an astonishingly good quantitative agreement
has been obtained. This is remarkable, especially if one considers the
fact that there is only very little room for ``fine-tuning'' of
parameters in the calculations that we have presented here. The
freedom we have to adjust the model is limited to the choice of the
initial random seed perturbation (which influences the relative
contribution of different low-order unstable modes in the post-shock
flow during the shock revival phase), and the parametrization of the
inner boundary condition which determines the (electron flavour)
neutrino flux from the neutron-star core, $L^{\rm core}$, and thereby
the explosion time scale and energy of the simulation
\citep[cf.][]{Scheck+06}. Note also that the core neutrino flux is
only a part of the total neutrino flux in our calculations. A
significant part of the total neutrino luminosity is contributed by
the neutrino emission of mass accreted onto the neutron star, which is
taken into account by our transport scheme (see
\citealt{Scheck+06}). The consistency of our models with the above
list therefore makes us confident that the long-standing mystery
concerning the origin of the major observational properties of SN
1987\,A may have found its solution.

\section{Conclusions}
\label{sect:conclusions}

The main conclusion of the present study is that if a \emph{low-mode}
hydrodynamic instability (more precisely, a quadrupole, $l=2$,
mode-dominated instability with a smaller contribution from a dipole,
$l=1$, mode) establishes during the first second of a supernova in a
standard $15~\Msol$ blue supergiant star (see
\citealt{Scheck+04,Scheck+06} for the requirements of this), then the
ensuing explosion possesses properties which are in remarkable
quantitative agreement with the observational data of SN 1987\,A.

Among other features, explosions of this type (with an energy of $\sim
2\times10^{51}$~erg for the progenitor model we used) exhibit final
iron-group velocities of up to 3300\,km/s, strong mixing at the He/H
composition interface, with hydrogen being mixed downward in velocity
space to only 500~km/s, and a final prolate anisotropy of the ejecta
with a major to minor axis ratio of about 1.6.

The success of low-mode explosions to reproduce these observations is
based on two effects: The long-wavelength pattern with which the
Rayleigh-Taylor unstable Si/O and O/He interfaces of the presupernova
are perturbed, and the initial global deformation of the shock, which
induces a strong \emph{Richtmyer-Meshkov instability} at the He/H
interface.

Compared to our previous models \citep{Kifonidis+03}, which were
dominated by high-mode neutrino-driven convection, we find that the
low-mode unstable models presented here show significantly (i.e. up to
40\%) higher initial metal clump velocities. These high velocities, in
turn, keep the timescale of clump propagation through the He core
short -- in particular shorter than the timescale of reverse shock
formation near the He/H interface. Hence the fastest clumps in the
low-mode models never have to interact with this reverse shock, and
are thus not slowed down, in marked contrast to the high-mode case
\citep{Kifonidis+03}.

The occurrence of a strong Richtmyer-Meshkov instability at the He/H
interface is of crucial importance for producing the strong inward
mixing of hydrogen and the global anisotropy that is present during
the late expansion of the ejecta. The early deposition of vorticity
into the interface layer by the initially aspherical shock (at $t
\approx 100$\,s) is responsible for the fact that a late-time
global anisotropy of the ejecta can develop (at $t \approx
10\,000$\,s), \emph{although} the shock itself has long become
spherical by that time.

We find that the large-scale anisotropy imprinted on the ejecta and
the early shock by low-mode hydrodynamic instabilities during the
shock revival phase, and the ensuing Richtmyer-Meshkov instability at
the He/H interface are all that is needed to reproduce the major
observational features of SN 1987\,A. In other words, mechanisms
like jet formation, anisotropic neutrino emission, and rotation of the
progenitor are not required to explain the observational
characteristics we have summarized in Sect.~\ref{sect:SN_1987A},
provided that due attention is paid to account for all hydrodynamic
instabilities that occur in the problem and for their mutual
interaction.

A coherent picture of supernova explosions, powered by the
neutrino-heating mechanism, thus seems to emerge, in which non-radial
instabilities of the accretion shock like the advective-acoustic
instability of \cite{Foglizzo_Tagger00} and \cite{Foglizzo02} (see in
particular \citealt{Foglizzo_Galetti03,Foglizzo+05,Ohnishi+05}), or the
stalled accretion shock instability (SASI) of \cite{Blondin+03} and
\cite{Blondin_Mezzacappa05}, may be of significant importance for
globally distorting the supernova shock during the shock revival phase
and for accelerating the neutron star to velocities in excess of
several hundred km/s \citep{Scheck+04,Scheck+06}. Here we have
demonstrated that such an early non-radial, low-mode shock instability
may also be of crucial importance for the observational appearance of
a supernova, because it can provide the shock anisotropy which is
responsible for the very strong Richtmyer-Meshkov instability that we
observe at the He/H interface. Judging from the results of our
simulations, this latter instability also deserves much more
appreciation by supernova modelers than it has received so far.

Although our simulations make use of a polar grid and the
assumption of axisymmetry, a fact which we critically assess in
Sect.~\ref{sect:grid_constraints}, the development of the discussed
low-mode instabilities and the corresponding explosion asymmetries is
neither a numerical artifact, nor a consequence of nonconvergence of
the simulations due to a lack of resolution, or even code
deficiencies. The variability of the early explosion asymmetries found
in the present study, and in the work of \cite{Scheck+06}, is the
result of an extremely nonlinear, chaotic growth of the initially
imposed small random seed perturbations. No numerical convergence of
simulations (in the mathematical sense) can be expected in such a
situation, especially for morphological aspects like the shock
deformation or the matter distribution in the unstable layer. Such
aspects differ even between runs of \emph{exactly} the same model, if
these runs are carried out on different computers with slightly
different 64 bit round-off behaviour. However, integral quantities
like the explosion timescale and energy, or the neutron star mass,
show little differences between simulations which make use of the same
boundary conditions, in spite of large variations of the explosion
geometry. Based on a huge set of models this is discussed in detail in
\cite{Scheck+06}. Therefore we do not expect that the findings
presented here will be challenged by future simulations with higher
grid resolution or different numerical schemes.

Furthermore, the occurrence and the final extent of the mixing due to
the Richtmyer-Meshkov instability is fairly robust against the range
of early explosion asymmetries that was explored in the present
paper. All three low-mode models that we have computed for the
\cite{WPE88} progenitor (i.e., Models b18a, b18b, and b23a) show a
comparable extent for the inward mixing of hydrogen and the outward
mixing of metals in mass (for a comparison between Models b18b and
b23a, see Fig.~\ref{fig:mixing}). A larger sample of simulations is,
however, required for a more reliable assessment of this issue.

The non-rotating models that we have presented here predict
non-rotating neutron stars (unless an off-center kick imparts also a
sizeable spin to the neutron star). This is in agreement with the fact
that to date no pulsar has been found in SN 1987\,A. However, this
point could be satisfied equally well by a rotating neutron star with
a weak surface magnetic field. Such an object would not produce
significant pulsar activity, and its thermal cooling signature falls
below the present detection limit. It remains to be seen in future
simulations, though, whether rotating models can be constructed that
are also able to reproduce the other observational items that we have
summarized above.

A pulsar would, of course, also be lacking if a black hole had formed
in SN 1987\,A. This option, however, is quite unlikely for an
$18-20~\Msol$ progenitor and with the core structure of current
progenitor models would require an unacceptably soft nuclear equation
of state, which seems to be excluded by the recent detection of a
neutron star with a well-measured mass of $2.1 \pm 0.2~\Msol$
\citep{Nice+05}.

While we are well aware of the fact that comparably well resolved
three-dimensional models and detailed lightcurves, spectra, and
polarization data still remain to be calculated, we think that already
the present results provide support to the hypothesis that SN 1987\,A
was an explosion which fits into the framework of the neutrino-driven
explosion paradigm, and that its understanding does not require the
assumption of (more) controversial physics.

In the future we plan to run long-time simulations for Type Ib,c
explosions, in which the He/H interface is absent and the large
anisotropy of the metal and/or helium core which we found in our
models will be exposed to the observer from the outset. We believe
that the global asymmetries imprinted on the ejecta by low-mode
hydrodynamic instabilities in the neutrino-heated postshock layer may
also offer an explanation of the large-scale anisotropy of the element
distribution visible in the Cassiopeia~A supernova remnant
\citep{Hwang+04}. Our results in Figs.~\ref{fig:massvelo_b23a} and
\ref{fig:massvelo_b18b}, and the results of Paper~I, suggest that in
this case the velocities of iron-group material can be expected to be
even higher for a given energy of the explosion.

\begin{appendix}

\section{The AUSM+ flux}
\label{sect:ausmp}

We consider the construction of a numerical flux function for the
system of Euler equations
\begin{equation}
{\partial \vec{U} \over \partial t} + 
{\partial \vec{F}(\vec{U}) \over \partial x} = 0,
\label{eq:Euler}
\end{equation}
where
\begin{equation}
\vec{U} = \left(
             \begin{matrix}
               \rho   \\
               \rho u \\
               \rho E
             \end{matrix}
          \right), \quad
\vec{F}(\vec{U}) = 
	     \left(
             \begin{matrix}
               \rho u  \\
               \rho u^2 + p \\
               \rho u H
             \end{matrix}
             \right),
\end{equation}
$\rho$, $u$, and $p$ have their usual meanings,
\begin{equation}
E = e + \frac{u^2}{2} = H - \frac{p}{\rho} 
\end{equation}
is the (specific) total energy, $e$ the internal energy, and $H$ the
total enthalpy.  We shall only summarize the calculation of the first
order accurate AUSM+ flux for 1D Cartesian geometry here. The
extension of the method to higher orders of accuracy, curvilinear
grids, and more than one spatial dimension is done using standard
techniques, (see e.g. \citealt{Colella_Woodward84} and
\citealt{Liou96}) and will not be explained in detail here.

The AUSM+ scheme of \cite{Liou96} is a sequel to the Advection
Upstream Splitting Method (AUSM) that was introduced by
\cite{Liou_Steffen93}. A central idea in these second generation flux
vector split schemes is that the advective and acoustic waves of the
system (\ref{eq:Euler}) describe two physically distinct processes and
should be treated as such by the numerical scheme. To achieve this,
the following ansatz is made in AUSM+ to calculate the numerical flux,
$\vec{f}_{j+1/2}$, at the interface $j+1/2$ between zones $j$ and
$j+1$
\begin{equation}
\vec{f}_{j+1/2} = m_{j+1/2} a_{j+1/2} \vec{\Phi}_{j+1/2} +
                  \left(
                    \begin{matrix}
	              0 \\
                      p_{j+1/2} \\
                      0
                    \end{matrix}
                  \right),
\label{eq:f_p_splitting}
\end{equation}
i.e. $\vec{f}_{j+1/2}$ is written as the sum of a convective and a
pressure term. In Eq.~(\ref{eq:f_p_splitting}) the quantities
$m_{j+1/2}$, $a_{j+1/2}$, $p_{j+1/2}$ and $\vec{\Phi}_{j+1/2}$ are
interface values of the Mach number, sound speed, pressure and the
modified state vector $\vec{\Phi} = \left(\rho,\rho u,\rho H
\right)^{\rm T}$, which contains the total enthalpy, $H$, instead of
the total energy, $E$.

The scheme proceeds by accounting for the effects of rightgoing and
leftgoing nonlinear waves (with eigenvalues $u \pm a$ of the flux
Jacobian $\partial \vec{F}(\vec{U})/ \partial \vec{U}$) by a proper
definition of the quantities $m_{j+1/2}$ and $p_{j+1/2}$. The effects
of the linearly degenerate field with eigenvalue $u$ are finally taken
care of by a simple upwind selection of $\vec{\Phi}_{j+1/2}$.  The
choice of $a_{j+1/2}$ mainly affects the resolution of shocks (and is
discussed in more detail below). The formulae for $m_{j+1/2}$,
$p_{j+1/2}$, $a_{j+1/2}$, and $\vec{\Phi}_{j+1/2}$ given by
\cite{Liou96} guarantee upwinding, an exact resolution of stationary
shocks and of both stationary and moving contact
discontinuities. Furthermore they ensure the positivity of the scheme,
and most important for the present application, the avoidance of the
``odd-even-decoupling'' and carbuncle phenomena in multidimensional
applications (see \citealt{Liou00} for details).

Liou's formula for the interface Mach number is
\begin{equation}
m_{j+1/2} = {\cal M}^{+}(M_{j}) + {\cal M}^{-}(M_{j+1}),
\end{equation} 
where the contributions of the rightgoing and leftgoing waves have
been denoted with superscripts ``$+$'' and ``$-$'',
respectively. Here
\begin{equation}
M_{j} = u_j/a_{j+1/2}, \quad M_{j+1} = u_{j+1}/a_{j+1/2}
\end{equation}
are left and right Mach numbers that are computed using the common
interface sound speed $a_{j+1/2}$. The split Mach numbers ${\cal
M}^{\pm}$ are given by the functions
\begin{equation}
{\cal M}^{\pm}(M) = 
\begin{cases} 
\frac{1}{2} \left(M\pm \left| M \right| \right), & 
\text{if} \left| M \right| \geq 1, \\
{\cal M}^{\pm}_{\beta}(M),               &
\text{otherwise},
\end{cases}
\end{equation}
where the ${\cal M}^{\pm}_{\beta}$ are defined as
\begin{equation}
{\cal M}^{\pm}_{\beta}(M) = \pm \frac{1}{4} \left(M \pm 1 \right)^2 
                            \pm \beta \left( M^2 - 1 \right)^2, 
                            \quad \beta=1/8
\end{equation}
(and we have corrected a misprint in the corresponding equation of
\citealt{Liou96}). Similarly, the interface pressure is defined as
\begin{equation}
p_{j+1/2} = {\cal P}^{+}(M_{j})p_j + {\cal P}^{-}(M_{j+1})p_{j+1},
\end{equation}
where the split pressures are given by
\begin{equation}
{\cal P}^{\pm}(M) = 
\begin{cases} 
\frac{1}{2} \left(1 \pm {\rm sign}(M) \right), & 
\text{if} \left| M \right| \geq 1, \\
{\cal P}^{\pm}_{\alpha}(M),               &
\text{otherwise},
\end{cases}
\end{equation}
with 
\begin{equation}
{\cal P}^{\pm}_{\alpha}(M) = \frac{1}{4} \left(M \pm 1 \right)^2
                             (2 \mp M)
                            \pm \alpha M \left( M^2 - 1 \right)^2, 
                            \quad \alpha=3/16.
\nonumber
\end{equation}

The final upwind selection of $\vec{\Phi}_{j+1/2}$ can be written
concisely in terms of the quantities
\begin{equation}
m^{\pm}_{j+1/2} = \frac{1}{2} \left( m_{j+1/2} \pm
                  \left| m_{j+1/2} \right| \right).
\end{equation}
With their help the numerical flux at interface $j+1/2$ can be expressed
as
\begin{equation}
\vec{f}_{j+1/2} = a_{j+1/2} \left\{ m^{+}_{j+1/2} 
                                  \left(
                                  \begin{matrix}
                                    \rho   \\
                                    \rho u \\
                                    \rho H
                                  \end{matrix}
                                  \right)_{j} +
                                   m^{-}_{j+1/2} 
                                  \left(
                                  \begin{matrix}
                                     \rho   \\
                                     \rho u \\
                                     \rho H
                                  \end{matrix}
                                  \right)_{j+1}
                            \right\} +
                                  \left(
                                  \begin{matrix}
	                            0 \\
                                    p_{j+1/2} \\
                                    0
                                  \end{matrix}
                                  \right).
\nonumber
\end{equation}
What remains to be specified is the interface sound speed $a_{j+1/2}$.
In principle the simple arithmetic average
\begin{equation}
a_{j+1/2} = \frac{1}{2} \left( a_{j} + a_{j+1} \right)
\end{equation}
may be taken at the expense of slightly smearing shock waves.  The
formula of \cite{Liou96}
\begin{equation}
a_{j+1/2} = \min{\left(\tilde a_j, \tilde a_{j+1}\right)},
\end{equation}
with
\begin{equation}
\tilde a = a_{*}^{2} / \max{ \left( a_{*}, \left| u \right| \right) },
\end{equation}
and the critical speed of sound, $a_{*}$, calculated via the
isoenergetic condition for an ideal gas
\begin{equation}
H = \frac{a^2}{\gamma - 1} + \frac{1}{2} u^2
  = \frac{(\gamma+1)a_{*}^{2}}{2(\gamma-1)},
\label{eq:astar}
\end{equation}
gives sharper shocks. In particular it allows for exact resolution of
stationary shocks (the generalization of (\ref{eq:astar}) to more
general equations of state, in which the ratio of specific heats,
$\gamma$, is not a constant, is straightforward). Moving shocks are
generally captured within one to two zones.

Numerical experiments confirm that the scheme is substantially less
viscous than older flux vector splittings. In fact we have found the
AUSM+ flux to be even less viscous than the Godunov flux (obtained
from the exact solution of the Riemann problem), which makes the
former a somewhat less robust building block than the latter if used
in higher order schemes to compute problems with complicated shock
interactions (as e.g. the well-known blastwaves problem of
\citealt{Woodward_Colella84}). For the present application, however,
in which we employed the AUSM+ flux only near a single strong shock
(where furthermore the flattening algorithm of PPM reduced the spatial
reconstruction to first order accuracy), the method worked well.

\end{appendix}

\begin{acknowledgements}
We are grateful to F.-K.~Thielemann for supplying us with his rate
library, and to F.-K.~Thielemann and R.~Hix for a copy of their
nucleosynthesis code, which turned out to be very helpful for coding
and testing our own routines. We would also like to thank S.~Bruenn
and S.~Woosley for providing us with the initial data, S. Blinnikov
and F.~X. Timmes for discussions, and the anonymous referee and
V.~Utrobin for their constructive comments which have helped us to
improve our paper. K.K., L.S. and H.-Th.J. were supported by the
Sonderforschungsbereich 375 on ``Astroparticle Physics'' of the
Deutsche Forschungsgemeinschaft. The work of T.P. was supported by the
US Department of Energy under Grant No. B523820 to the Center of
Astrophysical Thermonuclear Flashes at the University of Chicago. The
simulations were performed on the IBM p655 of the Max-Planck-Institut
f\"ur Astrophysik, on an SGI Altix 350 on loan to the ASC FLASH Center
by SGI, on the NEC SX-5/3C of the Rechenzentrum Garching (RZG), and on
the IBM p690 clusters of the RZG and the John von Neumann Institute
for Computing, J\"ulich. We are especially grateful to the last two
institutions for their support.
\end{acknowledgements}

\bibliography{paper}

\begin{thebibliography}{74}
\expandafter\ifx\csname natexlab\endcsname\relax\def\natexlab#1{#1}\fi

\bibitem[{Bader \& Deuflhard(1983)}]{Bader_Deuflhard83}
Bader, G. \& Deuflhard, P. 1983, Numer. Math., 41, 373

\bibitem[{{Blinnikov}(1999)}]{Blinnikov99}
{Blinnikov}, S.~I. 1999, Astronomy Letters, 25, 359

\bibitem[{{Blondin} \& {Mezzacappa}(2005)}]{Blondin_Mezzacappa05}
{Blondin}, J.~M. \& {Mezzacappa}, A. 2005, \apj, submitted, astro-ph/0507181

\bibitem[{{Blondin} {et~al.}(2003){Blondin}, {Mezzacappa}, \&
  {DeMarino}}]{Blondin+03}
{Blondin}, J.~M., {Mezzacappa}, A., \& {DeMarino}, C. 2003, \apj, 584, 971

\bibitem[{{Bouchet} {et~al.}(1991){Bouchet}, {Phillips}, {Suntzeff},
  {Gouiffes}, {Hanuschik}, \& {Wooden}}]{Bouchet+91}
{Bouchet}, P., {Phillips}, M.~M., {Suntzeff}, N.~B., {et~al.} 1991, \aap, 245,
  490

\bibitem[{Bruenn(1993)}]{Bruenn93}
Bruenn, S.~W. 1993, in Nuclear Physics in the Universe, ed. M.~W. Guidry \&
  M.~R. Strayer (Bristol: IOP), 31

\bibitem[{Bruenn {et~al.}(2006)Bruenn, Raley, \& Mezzacappa}]{Bruenn+06}
Bruenn, S.~W., Raley, E.~A., \& Mezzacappa, A. 2006, \apj, submitted,
  astro-ph/0404099

\bibitem[{{Buras} {et~al.}(2003){Buras}, {Rampp}, {Janka}, \&
  {Kifonidis}}]{Buras+03}
{Buras}, R., {Rampp}, M., {Janka}, H.-T., \& {Kifonidis}, K. 2003, \prl, 90,
  241101

\bibitem[{{Buras} {et~al.}(2006{\natexlab{a}}){Buras}, {Rampp}, {Janka}, \&
  {Kifonidis}}]{Buras+06}
---. 2006{\natexlab{a}}, \aap, 447, 1049

\bibitem[{{Buras} {et~al.}(2006{\natexlab{b}}){Buras}, {Rampp}, {Janka}, \&
  {Kifonidis}}]{Buras+06b}
---. 2006{\natexlab{b}}, \aap, submitted, astro-ph/0512189

\bibitem[{{Burrows} {et~al.}(1995){Burrows}, {Hayes}, \& {Fryxell}}]{BHF95}
{Burrows}, A., {Hayes}, J., \& {Fryxell}, B.~A. 1995, \apj, 450, 830

\bibitem[{{Chandrasekhar}(1961)}]{Chandra61}
{Chandrasekhar}, S. 1961, {Hydrodynamic and hydromagnetic stability} (Oxford:
  Clarendon)

\bibitem[{{Chevalier} \& {Soker}(1989)}]{Chevalier_Soker89}
{Chevalier}, R.~A. \& {Soker}, N. 1989, \apj, 341, 867

\bibitem[{Colella \& Woodward(1984)}]{Colella_Woodward84}
Colella, P. \& Woodward, P.~R. 1984, J. Comput. Phys., 54, 174

\bibitem[{{Colgan} {et~al.}(1994){Colgan}, {Haas}, {Erickson}, {Lord}, \&
  {Hollenbach}}]{Colgan+94}
{Colgan}, S. W.~J., {Haas}, M.~R., {Erickson}, E.~F., {Lord}, S.~D., \&
  {Hollenbach}, D.~J. 1994, \apj, 427, 874

\bibitem[{Einfeldt(1988)}]{Einfeldt88}
Einfeldt, B. 1988, SIAM J. Num. Anal., 25, 294

\bibitem[{{Fassia} \& {Meikle}(1999)}]{Fassia_Meikle99}
{Fassia}, A. \& {Meikle}, W. P.~S. 1999, \mnras, 302, 314

\bibitem[{{Foglizzo}(2002)}]{Foglizzo02}
{Foglizzo}, T. 2002, \aap, 392, 353

\bibitem[{{Foglizzo} \& {Galetti}(2003)}]{Foglizzo_Galetti03}
{Foglizzo}, T. \& {Galetti}, P. 2003, in Proc. of the workshop "3-D signatures
  in stellar explosions", June 10-13 2003, Austin, Texas, USA,
  astro--ph/0308534

\bibitem[{{Foglizzo} {et~al.}(2005){Foglizzo}, {Scheck}, \&
  {Janka}}]{Foglizzo+05}
{Foglizzo}, T., {Scheck}, L., \& {Janka}, H.-T. 2005, \apj, submitted,
  astro-ph/0507636

\bibitem[{{Foglizzo} \& {Tagger}(2000)}]{Foglizzo_Tagger00}
{Foglizzo}, T. \& {Tagger}, M. 2000, \aap, 363, 174

\bibitem[{{Fryer}(1999)}]{Fryer99}
{Fryer}, C.~L. 1999, \apj, 522, 413

\bibitem[{{Fryer} \& {Warren}(2002)}]{FW02}
{Fryer}, C.~L. \& {Warren}, M.~S. 2002, \apjl, 574, L65

\bibitem[{{Fryer} \& {Warren}(2004)}]{Fryer+04}
---. 2004, \apj, 601, 391

\bibitem[{{Haas} {et~al.}(1990){Haas}, {Erickson}, {Lord}, {Hollenbach},
  {Colgan}, \& {Burton}}]{Haas+90}
{Haas}, M.~R., {Erickson}, E.~F., {Lord}, S.~D., {et~al.} 1990, \apj, 360, 257

\bibitem[{{Hawley} \& {Zabusky}(1989)}]{Hawley_Zabusky89}
{Hawley}, J.~F. \& {Zabusky}, N.~J. 1989, \prl, 63, 1241

\bibitem[{{Herant}(1995)}]{Herant95}
{Herant}, M. 1995, Phys. Rep., 256, 117

\bibitem[{Herant \& Benz(1992)}]{Herant_Benz92}
Herant, M. \& Benz, W. 1992, \apj, 387, 294

\bibitem[{Herant {et~al.}(1994)Herant, Benz, Hix, Fryer, \& Colgate}]{HBFC94}
Herant, M., Benz, W., Hix, W.~R., Fryer, C.~L., \& Colgate, S.~A. 1994, \apj,
  435, 339

\bibitem[{{Hix} \& {Thielemann}(1996)}]{Hix_Thielemann96}
{Hix}, W.~R. \& {Thielemann}, F.-K. 1996, \apj, 460, 869

\bibitem[{{Hoffman} {et~al.}(1999){Hoffman}, {Woosley}, {Weaver}, {Rauscher},
  \& {Thielemann}}]{Hoffman+99}
{Hoffman}, R.~D., {Woosley}, S.~E., {Weaver}, T.~A., {Rauscher}, T., \&
  {Thielemann}, F.-K. 1999, \apj, 521, 735

\bibitem[{{Hungerford} {et~al.}(2005){Hungerford}, {Fryer}, \&
  {Rockefeller}}]{Hungerford+05}
{Hungerford}, A.~L., {Fryer}, C.~L., \& {Rockefeller}, G. 2005, \apj, 635, 487

\bibitem[{{Hungerford} {et~al.}(2003){Hungerford}, {Fryer}, \&
  {Warren}}]{Hungerford+03}
{Hungerford}, A.~L., {Fryer}, C.~L., \& {Warren}, M.~S. 2003, \apj, 594, 390

\bibitem[{{Hwang} {et~al.}(2004){Hwang}, {Laming}, {Badenes}, {Berendse},
  {Blondin}, {Cioffi}, {DeLaney}, {Dewey}, {Fesen}, {Flanagan}, {Fryer},
  {Ghavamian}, {Hughes}, {Morse}, {Plucinsky}, {Petre}, {Pohl}, {Rudnick},
  {Sankrit}, {Slane}, {Smith}, {Vink}, \& {Warren}}]{Hwang+04}
{Hwang}, U., {Laming}, J.~M., {Badenes}, C., {et~al.} 2004, \apjl, 615, L117

\bibitem[{Janka \& M\"uller(1996)}]{JM96}
Janka, H.-T. \& M\"uller, E. 1996, \aap, 306, 167

\bibitem[{{Jennings} {et~al.}(1993){Jennings}, {Boyle}, {Wiedemann}, \&
  {Moseley}}]{Jennings+93}
{Jennings}, D.~E., {Boyle}, R.~J., {Wiedemann}, G.~R., \& {Moseley}, S.~H.
  1993, \apj, 408, 277

\bibitem[{{Kane} {et~al.}(2000){Kane}, {Arnett}, {Remington}, {Glendinning},
  {Baz\a'an}, {M\"uller}, {Fryxell}, \& {Teyssier}}]{Kane+00}
{Kane}, J., {Arnett}, W.~D., {Remington}, B.~A., {et~al.} 2000, \apj, 528, 989

\bibitem[{{Kifonidis} {et~al.}(2000){Kifonidis}, {Plewa}, {Janka}, \&
  {M\"uller}}]{Kifonidis+00}
{Kifonidis}, K., {Plewa}, T., {Janka}, H.-T., \& {M\"uller}, E. 2000, \apjl,
  531, L123

\bibitem[{{Kifonidis} {et~al.}(2003){Kifonidis}, {Plewa}, {Janka}, \&
  {M\"uller}}]{Kifonidis+03}
---. 2003, \aap, 408, 621 (Paper~I)

\bibitem[{{Kozma} \& {Fransson}(1998)}]{Kozma_Fransson98}
{Kozma}, C. \& {Fransson}, C. 1998, \apj, 497, 431

\bibitem[{Liou(1996)}]{Liou96}
Liou, M.-S. 1996, J. Comput. Phys., 129, 364

\bibitem[{Liou(2000)}]{Liou00}
---. 2000, J. Comput. Phys., 160, 623

\bibitem[{{Liou} \& {Steffen}(1993)}]{Liou_Steffen93}
{Liou}, M.-S. \& {Steffen}, C.~J. 1993, J. Comput. Phys., 107, 23

\bibitem[{{McCray}(1993)}]{McCray93}
{McCray}, R. 1993, \araa, 31, 175

\bibitem[{Meshkov(1969)}]{Meshkov_69}
Meshkov, E.~E. 1969, Izv. Akad. Nauk SSSR, Mekh. Zhidk. Gaza, 4, 151

\bibitem[{{Miles}(2004)}]{Miles04}
{Miles}, A.~R. 2004, Physics of Plasmas, 11, 5140

\bibitem[{{Mitchell} {et~al.}(2001){Mitchell}, {Baron}, {Branch}, {Lundqvist},
  {Blinnikov}, {Hauschildt}, \& {Pun}}]{Mitchell+01}
{Mitchell}, R.~C., {Baron}, E., {Branch}, D., {et~al.} 2001, \apj, 556, 979

\bibitem[{M{\"u}ller(1986)}]{Mueller86}
M{\"u}ller, E. 1986, \aap, 162, 103

\bibitem[{{Nagataki}(2000)}]{Nagataki00}
{Nagataki}, S. 2000, \apjs, 127, 141

\bibitem[{Nice {et~al.}(2005)Nice, Splaver, Stairs, Loehmer, Jessner, Kramer,
  \& Cordes}]{Nice+05}
Nice, D., Splaver, E.~M., Stairs, I.~H., {et~al.} 2005, \apj, {submitted,
  astro-ph/0508050}

\bibitem[{Ohnishi {et~al.}(2005)Ohnishi, Kotake, \& Yamada}]{Ohnishi+05}
Ohnishi, N., Kotake, K., \& Yamada, S. 2005, \apj, submitted, astro-ph/0509765

\bibitem[{{Phillips} {et~al.}(1990){Phillips}, {Hamuy}, {Heathcote},
  {Suntzeff}, \& {Kirhakos}}]{Phillips+90}
{Phillips}, M.~M., {Hamuy}, M., {Heathcote}, S.~R., {Suntzeff}, N.~B., \&
  {Kirhakos}, S. 1990, \aj, 99, 1133

\bibitem[{Press {et~al.}(1992)Press, Teukolsky, Vetterling, \&
  Flannery}]{Press+92}
Press, W.~H., Teukolsky, S.~A., Vetterling, W.~T., \& Flannery, B.~P. 1992,
  Numerical Recipes in FORTRAN, The Art of Scientific Computing, Second Edition
  (Cambridge: Cambridge University Press)

\bibitem[{Richtmyer(1960)}]{Richtmyer_60}
Richtmyer, R.~D. 1960, Commun. Pure Appl. Math., 13, 297

\bibitem[{Ronchi {et~al.}(1996)Ronchi, Iacono, \& Paolucci}]{Ronchi+96}
Ronchi, C., Iacono, R., \& Paolucci, P.~S. 1996, J. Comput. Phys., 124, 93

\bibitem[{{Scheck} {et~al.}(2006){Scheck}, {Kifonidis}, {Janka}, \&
  {M{\"u}ller}}]{Scheck+06}
{Scheck}, L., {Kifonidis}, K., {Janka}, H.-T., \& {M{\"u}ller}, E. 2006, \aap,
  submitted, astro-ph/0601302

\bibitem[{{Scheck} {et~al.}(2004){Scheck}, {Plewa}, {Janka}, {Kifonidis}, \&
  {M{\"u}ller}}]{Scheck+04}
{Scheck}, L., {Plewa}, T., {Janka}, H.-T., {Kifonidis}, K., \& {M{\"u}ller}, E.
  2004, \prl, 92, 011103

\bibitem[{{Shigeyama} \& {Nomoto}(1990)}]{Shigeyama_Nomoto90}
{Shigeyama}, T. \& {Nomoto}, K.~I. 1990, \apj, 360, 242

\bibitem[{Socrates {et~al.}(2005)Socrates, Blaes, Hungerford, \&
  {Fryer}}]{Socrates+05}
Socrates, A., Blaes, O., Hungerford, A., \& {Fryer}, C. 2005, \apj, {submitted,
  astro-ph/0412144}

\bibitem[{{Spyromilio} {et~al.}(1990){Spyromilio}, {Meikle}, \&
  {Allen}}]{Spyromilio+90}
{Spyromilio}, J., {Meikle}, W.~P.~S., \& {Allen}, D.~A. 1990, \mnras, 242, 669

\bibitem[{{Suntzeff} \& {Bouchet}(1990)}]{Suntzeff_Bouchet90}
{Suntzeff}, N.~B. \& {Bouchet}, P. 1990, \aj, 99, 650

\bibitem[{Thielemann {et~al.}(1996)Thielemann, Nomoto, \& Hashimoto}]{TNH96}
Thielemann, F.-K., Nomoto, K.~I., \& Hashimoto, M. 1996, \apj, 460, 408

\bibitem[{{Thompson} {et~al.}(2005){Thompson}, {Quataert}, \&
  {Burrows}}]{Thompson+05}
{Thompson}, T.~A., {Quataert}, E., \& {Burrows}, A. 2005, \apj, 620, 861

\bibitem[{{Timmes}(1999)}]{Timmes99}
{Timmes}, F.~X. 1999, \apjs, 124, 241

\bibitem[{{Timmes} \& {Swesty}(2000)}]{Timmes_Swesty00}
{Timmes}, F.~X. \& {Swesty}, F.~D. 2000, \apjs, 126, 501

\bibitem[{{Utrobin}(2004)}]{Utrobin04}
{Utrobin}, V.~P. 2004, Astronomy Letters, 30, 293

\bibitem[{{Utrobin} \& {Chugai}(2005)}]{Utrobin_Chugai05}
{Utrobin}, V.~P. \& {Chugai}, N.~N. 2005, \aap, 441, 271

\bibitem[{{Utrobin} {et~al.}(1995){Utrobin}, {Chugai}, \&
  {Andronova}}]{Utrobin+95}
{Utrobin}, V.~P., {Chugai}, N.~N., \& {Andronova}, A.~A. 1995, \aap, 295, 129

\bibitem[{{Wang} {et~al.}(2002){Wang}, {Wheeler}, {H{\" o}flich}, {Khokhlov},
  {Baade}, {Branch}, {Challis}, {Filippenko}, {Fransson}, {Garnavich},
  {Kirshner}, {Lundqvist}, {McCray}, {Panagia}, {Pun}, {Phillips}, {Sonneborn},
  \& {Suntzeff}}]{Wang+02}
{Wang}, L., {Wheeler}, J.~C., {H{\" o}flich}, P., {et~al.} 2002, \apj, 579, 671

\bibitem[{{Wilson} {et~al.}(2005){Wilson}, {Mathews}, \& {Dalhed}}]{Wilson+05}
{Wilson}, J.~R., {Mathews}, G.~J., \& {Dalhed}, H.~E. 2005, \apj, 628, 335

\bibitem[{Woodward \& Colella(1984)}]{Woodward_Colella84}
Woodward, P.~R. \& Colella, P. 1984, J. Comput. Phys., 54, 115

\bibitem[{{Woosley}(1988)}]{Woosley88}
{Woosley}, S.~E. 1988, \apj, 330, 218

\bibitem[{{Woosley} {et~al.}(1988){Woosley}, {Pinto}, \& {Ensman}}]{WPE88}
{Woosley}, S.~E., {Pinto}, P.~A., \& {Ensman}, L. 1988, \apj, 324, 466

\bibitem[{{Yamada} \& {Sato}(1991)}]{Yamada_Sato91}
{Yamada}, S. \& {Sato}, K. 1991, \apj, 382, 594

\end{thebibliography}

\end{document}